\documentclass[journal]{IEEEtran}

\usepackage{cite}
\usepackage{amsmath,amssymb,amsfonts}
\usepackage{algorithmic}
\usepackage{graphicx}
\usepackage{textcomp}
\usepackage{amssymb}
\usepackage{latexsym}
\usepackage{rotating}
\usepackage{color, colortbl}
\usepackage{array, caption, floatrow, tabularx, makecell, booktabs}%
\usepackage{layouts}
\usepackage{graphicx}
\usepackage{siunitx}
\usepackage{url}
\usepackage[dvipsnames]{xcolor}
\usepackage{hyperref}
\usepackage[normalem]{ulem}
\usepackage{fontawesome}

\definecolor{ConvexAdam}{HTML}{82538B}
\definecolor{LapIRN}{HTML}{3E6D9B}
\definecolor{MEVIS}{HTML}{5FB346}
\definecolor{corrField}{HTML}{C74342}
\definecolor{NiftyReg}{HTML}{929292}
\definecolor{PDD-Net}{HTML}{A9A9CC}
\definecolor{PIMed}{HTML}{EBAF56}
\definecolor{Gunnarsson}{HTML}{A9BCCC}
\definecolor{IWM}{HTML}{CCC8A9}
\definecolor{Estienne}{HTML}{B7CCA9}
\definecolor{Joutard}{HTML}{F8DF81}
\definecolor{Driver}{HTML}{B8CC90}
\definecolor{Jaouen}{HTML}{A9CBCC}
\definecolor{Winter}{HTML}{EBB4B3}
\definecolor{Imperial}{HTML}{CCA9C7}
\definecolor{Brudfors}{HTML}{CCBAA9}
\definecolor{3Idiots}{HTML}{EBCA9A}
\definecolor{Thorley}{HTML}{CCA9A9}
\definecolor{Bailiang}{HTML}{92ACB2}
\definecolor{Fourcade}{HTML}{FFCCB6}
\definecolor{Lifshitz}{HTML}{B9BECC}

\definecolor{gold}{HTML}{D4AF37}
\definecolor{silver}{HTML}{C0C0C0}
\definecolor{bronze}{HTML}{9F7A34}

\definecolor{newcolor}{rgb}{.8,.349,.1}
\definecolor{Gray}{gray}{0.9}

\newcommand\removed[1]{}
\newcommand{\revised}[1]{\textcolor{black}{#1}}

\newcolumntype{D}[1]{>{\centering\arraybackslash}p{#1}}
\newcommand{\stimes}{{\times}}
\newcommand{\sapprox}{{\sim}}
\newcommand\sbullet[1][2.5]{\mathbin{\vcenter{\hbox{\scalebox{#1}{$\bullet$}}}}}
\newcolumntype{C}{>{\centering\arraybackslash}X}
\newcommand\x[1][1.25]{$\mathbin{\vcenter{\hbox{\scalebox{#1}{$\bullet$}}}}$}
\newcommand\y[1][0.75]{$\mathbin{\vcenter{\hbox{\scalebox{#1}{$\blacksquare$}}}}$}
\newcommand\z[1][1.25]{$\mathbin{\vcenter{\hbox{\scalebox{#1}{$\circ$}}}}$}
\newcolumntype{?}{!{\vrule width 1pt}}

\begin{document}
\title{Learn2Reg: comprehensive multi-task medical image registration challenge, dataset and evaluation in the era of deep learning}
\author{Alessa~Hering*,
        Lasse~Hansen*$^\dagger$,
        Tony~C.~W.~Mok,
        Albert~C.~S.~Chung,
        Hanna~Siebert,
        Stephanie~H\"ager,
        Annkristin~Lange,
        Sven~Kuckertz,
        Stefan~Heldmann,
        Wei~Shao,
        Sulaiman~Vesal,
        Mirabela~Rusu,
        Geoffrey~Sonn,
        Th\'{e}o~Estienne,
        Maria~Vakalopoulou,
        Luyi~Han,
        Yunzhi~Huang,
        Pew-Thian Yap,
        Mikael~Brudfors,
        Ya\"el~ Balbastre,
        Samuel~Joutard,
        Marc~Modat,
        Gal~Lifshitz,
        Dan~Raviv,
        Jinxin~Lv,
        Qiang~Li,
        Vincent~Jaouen,
        Dimitris~Visvikis,
        Constance~Fourcade,
        Mathieu~Rubeaux,
        Wentao~Pan,
        Zhe~Xu,
        Bailiang~Jian,
        Francesca~De~Benetti,
        Marek~Wodzinski,
        Niklas~Gunnarsson,
        Jens Sjölund,
        Daniel Grzech,
        Huaqi~Qiu,
        Zeju~Li, 
        \revised{Alexander~Thorley},
        \revised{Jinming~Duan},
        Christoph~Großbr\"ohmer,
        Andrew Hoopes,
        Ingerid~Reinertsen,
        Yiming~Xiao,
        Bennett~Landman,
        Yuankai~Huo,
        Keelin~Murphy,
        Nikolas Lessmann,
        Bram~van~Ginneken,
        Adrian~V.~Dalca,
        Mattias~P.~Heinrich

\thanks{*Alessa Hering and Lasse Hansen contributed equally to this work.}
\thanks{${}^\dagger$Corresponding author: \url{hansen@imi.uni-luebeck.de}, Institute of Medical Informatics, Universität zu Lübeck, Ratzeburger Allee 160, 23562 Lübeck, Germany}
\thanks{Author affiliations are listed at the end of the paper.}}

\maketitle

\begin{abstract}
Image registration is a fundamental medical image analysis task, and a wide variety of approaches have been proposed. However, only a few studies have comprehensively compared medical image registration approaches on a wide range of clinically relevant tasks. This limits the development of registration methods, the adoption of research advances into practice, and a fair benchmark across competing approaches. The Learn2Reg challenge addresses these limitations by providing a multi-task medical image registration data set for comprehensive characterisation of deformable registration algorithms. A continuous evaluation will be possible at \url{https://learn2reg.grand-challenge.org}. Learn2Reg covers a wide range of anatomies (brain, abdomen, and thorax), modalities (ultrasound, CT, MR), availability of annotations, as well as intra- and inter-patient registration evaluation. We established an easily accessible framework for training and validation of 3D registration methods, which enabled the compilation of results of over 65 individual method submissions from more than 20 unique teams. We used a complementary set of metrics, including robustness, accuracy, plausibility, and runtime, enabling unique insight into the current state-of-the-art of medical image registration. This paper describes datasets, tasks, evaluation methods and results of the challenge, as well as results of further analysis of transferability to new datasets, the importance of label supervision, and resulting bias. While no single approach worked best across all tasks, many methodological aspects could be identified that push the performance of medical image registration to new state-of-the-art performance. Furthermore, we demystified the common belief that conventional registration methods have to be much slower than deep-learning-based methods.
\end{abstract}


\section{Introduction}
\label{sec:introduction}
\IEEEPARstart{I}{mage} registration is a fundamental task in medical image analysis and has been an active field of research for decades~\cite{maintz1998surveyIR,SotirasDavatzikosParagios2013,viergever2016survey,haskins2020deep}. Most studies that compared registration methods were focused on specific tasks or algorithmic aspects, and did not comprehensively characterise current approaches. With the recent success of deep learning strategies in image analysis, the degree and dependency of algorithms on (partially) labelled training data is often a crucial aspect in current research. The Learn2Reg challenge aims to gain insight into which methodological components and supervision strategies are best suited for a wide range of clinically useful 3D image registration tasks, and sets a new benchmark to evaluate and distinguish strengths and weaknesses of task-tailored solutions.
Learn2Reg covers a wide range of anatomies (brain, abdomen and thorax), modalities (ultrasound, CT, MR) and auxiliary  annotations (e.g. segmentation, keypoints). The challenge also includes both intra- and inter-patient registration tasks. Due to this broad range, it serves as a unique benchmark to evaluate the current state-of-the-art with respect to various qualities of registration algorithms: accuracy, robustness, plausibility and speed. Furthermore, no other medical image registration challenge has thoroughly analysed the benefits and shortcomings of learning- and optimisation-based strategies. To engage a wider participation from new research groups, Learn2Reg removes entry barriers by providing pre-processed and pre-aligned images with additional annotations, as well as evaluation scripts and code for all evaluation metrics. 

This overview ranks and scores results from over 65 entries from more than 20 teams throughout 2020 and 2021. We perform additional experiments to analyse the robustness towards cross-dataset transfer, the influence of the bias induced by only labelling certain anatomical regions, and direct comparisons of the supervision level of selected methods.

\subsection{Related Work}
In the following a brief overview of important related work on comparing (bio)-medical image registration, and its fundamental methodological choices that differentiate the wide range of metrics, optimisation, and supervision is given. General guidelines for setting up a fair and unbiased challenge have been recently thoroughly discussed in literature~\cite{maier2020bias}. These criteria were adhered to in Learn2Reg and externally reviewed and confirmed by the MICCAI challenge team.

\paragraph*{\textbf{Challenges}} There have previously been four prominent challenges for medical image registration. Three challenges focused on a single task: EMPIRE10 (lung CT)~\cite{murphy2011empire10}, CuRIOUS (intra-operative US and MR)~\cite{xiao2019evaluation}, and ANHIR (histology)~\cite{borovec2020anhir}. Each attracted at least ten participating teams and used various metrics for quantifying the performance. The EMPIRE10 challenge provided the most comprehensive evaluation including distances of manual landmark pairs, fissure segmentations, and Jacobian determinant values of the deformation field. This challenge also required (original) participants to perform live registrations during the MICCAI workshop in Beijing and therefore employed a time constraint on the computations. The Continuous Registration Challenge~\cite{marstal2019continuous} co-organised with WBIR 2018 aimed at combining multiple tasks from previous benchmarks (lung CT and inter-patient brain MR). It addressed assessing registration quality as a service but is limited to algorithms that can be incorporated into the SuperElastix framework and therefore had limited participation.

\paragraph*{\textbf{Benchmark Papers}} Several papers have compared multiple registration algorithms for a given dataset. In contrast to challenges, these benchmark papers did not have an open workshop format that enabled wide-spread participation. Nevertheless, their findings provided meaningful insights. Starting from RIRE~\cite{west1997comparison}, which compared rigid-body alignment of head MR (T1, T2), PET and CT, there have been several brain registration benchmarks - most notably the evaluation of 14 nonlinear iterative registration algorithms~\cite{klein2009evaluation}. Fewer studies analysed  abdominal registration, and included the evaluation of six affine and non-linear algorithms on inter-patient registration of the "beyond the cranial vault" dataset~\cite{xu2016evaluation}. This study revealed large performance gaps and motivated our inclusion of this dataset to study the potential benefit of supervised (learning-based) algorithms. The DIR-Lab datasets~\cite{castillo2009framework} have been widely used to benchmark intra-patient CT lung motion estimation and provide a leaderboard for state-of-the-art comparison. All landmarks are publicly available, which makes the dataset prone to overfitting on the test data.

\paragraph*{\textbf{Survey Papers and Baseline Methods}}
Surveys on conventional medical image registration~\cite{SotirasDavatzikosParagios2013,viergever2016survey} have comprehensively reviewed typical categories of approaches including similarity metric, regulariser, and optimiser criteria. Due to the strong increase in the number of deep-learning-based registration paper in the last few years, several new surveys have been published (e.g. ~\cite{haskins2020deep}) extending the typical categories with deep-learning specific categories like supervision-type and network architecture. Moreover, the training data and thus the registered body region and image modality are more important for deep-learning-based methods and get more into the focus of those survey papers. While few papers have evaluated their proposed registration method on more than two different registration tasks, there is a variety of public methods SyN~\cite{avants2008symmetric}, Elastix~\cite{klein2009elastix}, NiftyReg~\cite{modat2010fast} and deeds~\cite{heinrich2013mrf}, and Voxelmorph~\cite{dalca2019unsupervised} that are commonly used as baseline or comparison methods. When comparing only among deep-learning based methods simply re-training specific architectures on new data may be insufficient. Hence the use of a challenge benchmark that incorporates several generally applicable baselines is essential for a fair evaluation.

\subsection{Contributions}
Learn2Reg provides both datasets and an easily accessible benchmark for the first comprehensive evaluation of a wide-range of methods for inter- and intra-patient, mono- and multimodal medical registration.  We introduce a complementary set of metrics, including robustness, accuracy, plausibility and speed, that follows the principles defined by the BIAS group~\cite{maier2020bias} and could become an important data set collection for comparing new algorithms. Further analysis of label bias (for supervised methods), domain transfer and statistical testing of significant differences across algorithms and types of methods highlight the complementary strength and weaknesses of learning vs. non-learning-based approaches.

\section{Material and Methods}
\subsection{Challenge Organisation} 

The Learn2Reg challenge is organised by Alessa Hering, Lasse Hansen, Adrian Dalca  and Mattias Heinrich and is associated with MICCAI 2020 and 2021. The following tasks were included in 2020: CuRIOUS, Hippocampus MR, Abdomen CT-CT and Lung CT. In 2021, Abdomen MR-CT and OASIS were newly introduced and the Lung CT task was continued. The Learn2Reg challenge consisted of two phases (mainly organised on grand-challenge.org).
\begin{itemize}
    \item Phase 1 - Validation Phase: \revised{The participants downloaded the training and validation datasets and trained a registration network or tuned hyperparameters on them. }\removed{The participants downloaded the training and validation scan pairs for each task described in section \ref{sec:tasks}. The participants trained a registration network or tuned hyperparameters on the training scan pairs in their own facilities. The developed algorithms were used to register the scan pairs of the validation dataset.} The calculated displacement fields on the validation dataset were submitted and evaluated using grand-challenge.org. Challenge participants were allowed to create five submissions per day to this phase.
    
    \item Phase 2 - Test phase: Within one week after the test data release, the participants had to send either the generated displacement fields to the organisers or a Docker container containing the algorithm. A Docker submission was preferred and made more attractive by evaluating the runtime of the algorithm.
\end{itemize}
Members of the organisers' institutes could participate in the challenge having the same data access as any other participant. However, they were not eligible for awards. A continuous evaluation for test data will be possible at grand-challenge.org\footnote{\url{https://learn2reg.grand-challenge.org}\\\hspace*{11.5pt}\url{https://learn2reg-test.grand-challenge.org}}. All methods that solve at least four of the six tasks are included into the overall ranking of this paper. To remove entry barriers for new participants with expertise in deep learning but not necessarily registration, the organisers provided pre-preprocessed data. A detailed description of the used preprocessing is given in section \ref{sec:tasks}. Furthermore, the evaluation code for voxel displacement fields as well as an example Docker container submissions were provided. All additional resources can be found at the Learn2Reg repository\footnote{\url{https://github.com/MDL-UzL/L2R}}.
\subsection{Tasks}
\label{sec:tasks}
\setlength{\tabcolsep}{2pt}
\renewcommand{\arraystretch}{0.75}
\begin{table*}[!h]
    \centering
    \begin{tabular}{rD{2.25cm}D{2.25cm}D{2.25cm}D{2.25cm}D{2.25cm}D{2.25cm}}
        \toprule[1.5pt]
        & \multicolumn{2}{c}{CuRIOUS} & \multicolumn{2}{c}{Hippocampus MR} & \multicolumn{2}{c}{Abdomen CT-CT} \\
          \cmidrule(lr){2-3}            \cmidrule(lr){4-5}            \cmidrule(lr){6-7}
        & Fixed & Moving              & Fixed & Moving              & Fixed & Moving \\
        & \includegraphics[width=2cm]{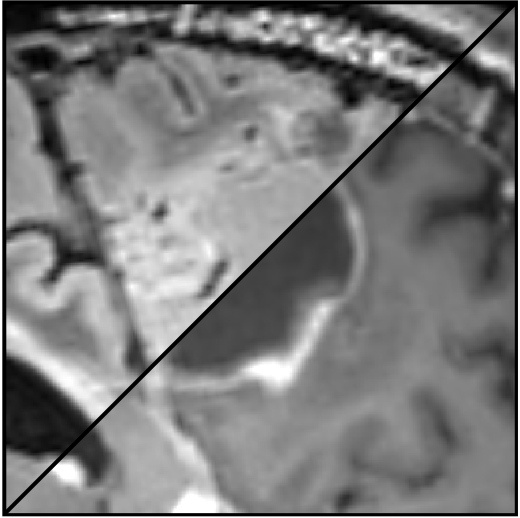} & \includegraphics[width=2cm]{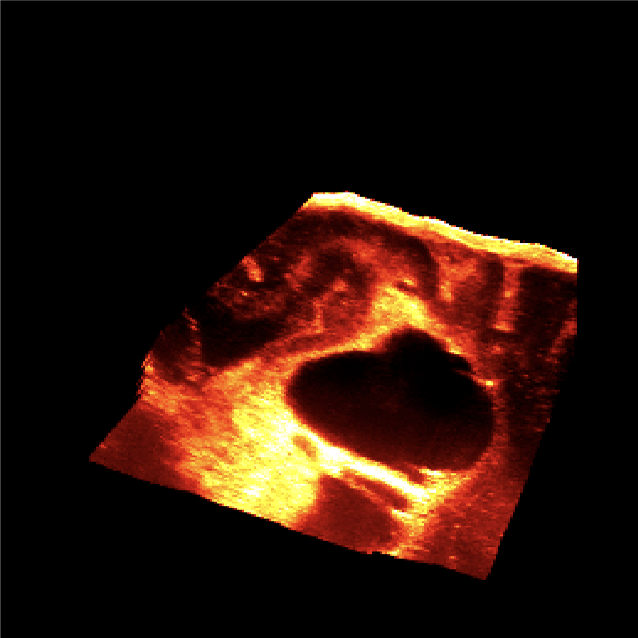} &
        \includegraphics[width=2cm]{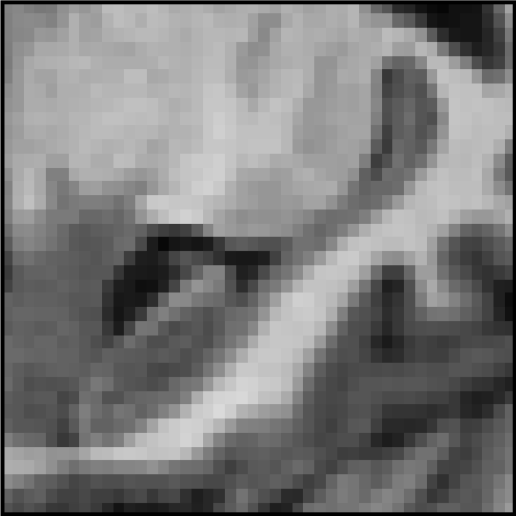} & \includegraphics[width=2cm]{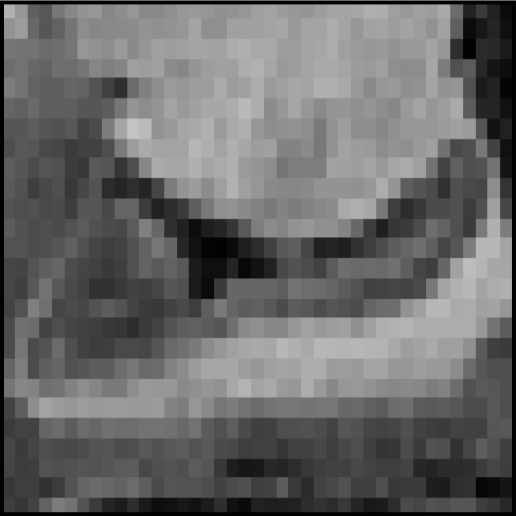} & \includegraphics[width=2cm]{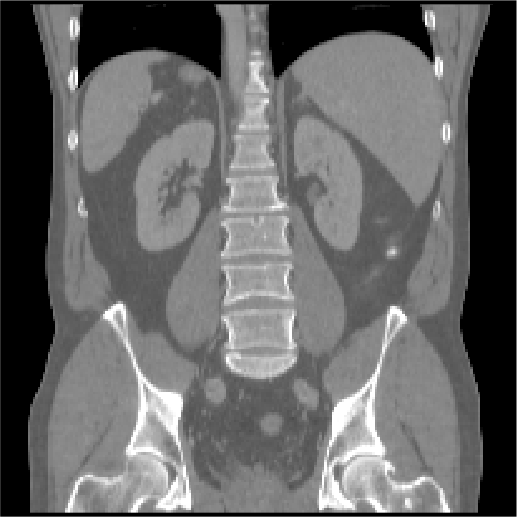} & \includegraphics[width=2cm]{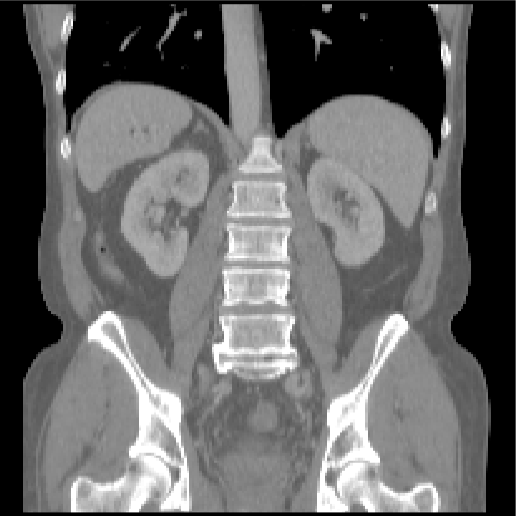}   \\
        \midrule
        Modalities & \multicolumn{2}{c}{MR T1w \& FLAIR/US} & \multicolumn{2}{c}{MR T1w/MR T1w} & \multicolumn{2}{c}{CT/CT}        \\[0.1cm]
        Intra-/Inter-patient     & \multicolumn{2}{c}{Intra-patient}         & \multicolumn{2}{c}{Inter-patient} & \multicolumn{2}{c}{Inter-patient} \\[0.1cm]
        Resolution & \multicolumn{2}{c}{$256\stimes256\stimes288$}  & \multicolumn{2}{c}{$64\stimes64\stimes64$} & \multicolumn{2}{c}{$192\stimes160\stimes256$} \\[0.1cm]
        Voxel size & \multicolumn{2}{c}{$\sapprox0.5\stimes0.5\stimes0.5\SI{}{\milli\meter}$} & \multicolumn{2}{c}{$1\stimes1\stimes1\SI{}{\milli\meter}$}       & \multicolumn{2}{c}{$2\stimes2\stimes2\SI{}{\milli\meter}$} \\[0.1cm]
        Cases (Train/Test) & \multicolumn{2}{c}{22/10}  & \multicolumn{2}{c}{263/131} & \multicolumn{2}{c}{30/20} \\[0.1cm]
        Preprocessing & \multicolumn{2}{c}{resample} & \multicolumn{2}{c}{crop/pad/resample} & \multicolumn{2}{c}{canonical affine pre-align} \\[0.05cm]
        & & & & & \multicolumn{2}{c}{crop/pad/resample} \\[0.1cm]
        Annotations (Train/Test) & \multicolumn{2}{c}{--/9-18 landmarks/case} & \multicolumn{2}{c}{2/2 anatomical labels} & \multicolumn{2}{c}{13/13 anatomical labels} \\[0.1cm]
        Additional data &  &  &  \\[0.1cm]
        Metrics & \multicolumn{2}{c}{TRE/TRE30} & \multicolumn{2}{c}{DSC/DSC30/HD95} & \multicolumn{2}{c}{DSC/DSC30/HD95} \\[0.05cm]
        & \multicolumn{2}{c}{SDlogJ/RT} & \multicolumn{2}{c}{SDlogJ/RT} & \multicolumn{2}{c}{SDlogJ/RT}\\[0.1cm]
        Challenges & \multicolumn{2}{l}{\textcolor{white}{$\sbullet$}\textcolor{OliveGreen}{$\sbullet$}\textcolor{MidnightBlue}{$\sbullet$}\textcolor{Thistle}{$\sbullet$}\textcolor{Goldenrod}{$\sbullet$}\textcolor{black}{$\sbullet$}\textcolor{Turquoise}{$\sbullet$}}  & \multicolumn{2}{l}{\textcolor{white}{$\sbullet$}\textcolor{RedOrange}{$\sbullet$}\textcolor{MidnightBlue}{$\sbullet$}\textcolor{Maroon}{$\sbullet$}} & \multicolumn{2}{l}{\textcolor{white}{$\sbullet$}\textcolor{Thistle}{$\sbullet$}\textcolor{Goldenrod}{$\sbullet$}} \\[0.25cm]
        & \multicolumn{2}{c}{Abdomen MR-CT} & \multicolumn{2}{c}{OASIS} & \multicolumn{2}{c}{Lung CT} \\
          \cmidrule(lr){2-3}            \cmidrule(lr){4-5}            \cmidrule(lr){6-7}
        & Fixed & Moving              & Fixed & Moving              & Fixed & Moving \\
        & \includegraphics[width=2cm]{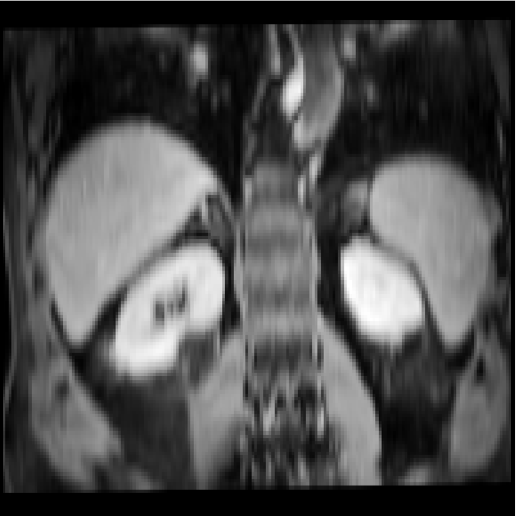} & \includegraphics[width=2cm]{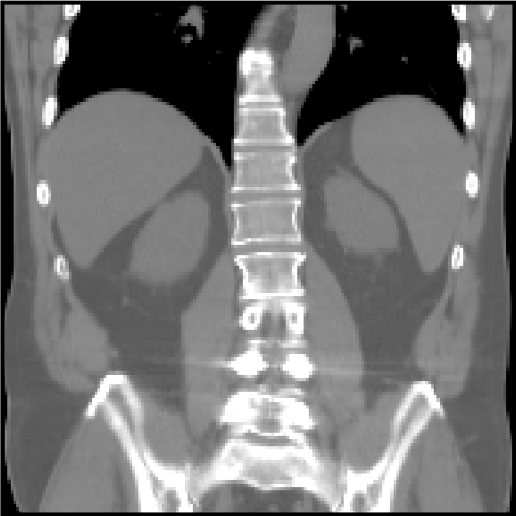} & \includegraphics[width=2cm]{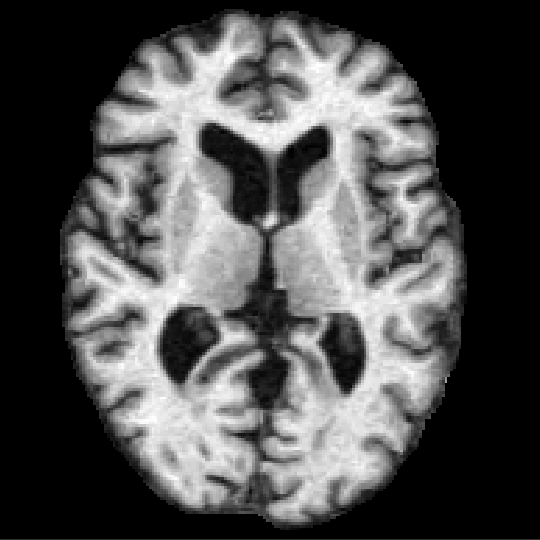} & \includegraphics[width=2cm]{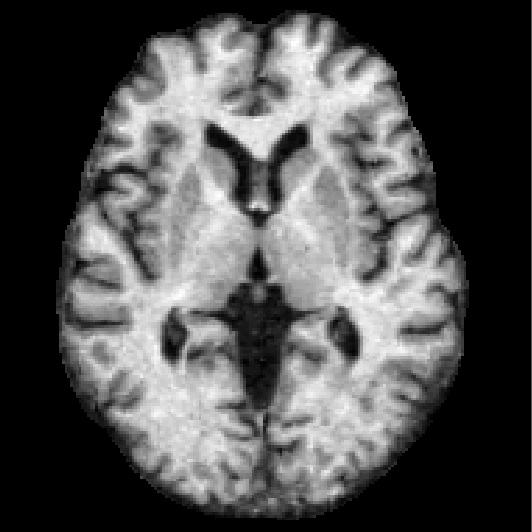} & \includegraphics[width=2cm]{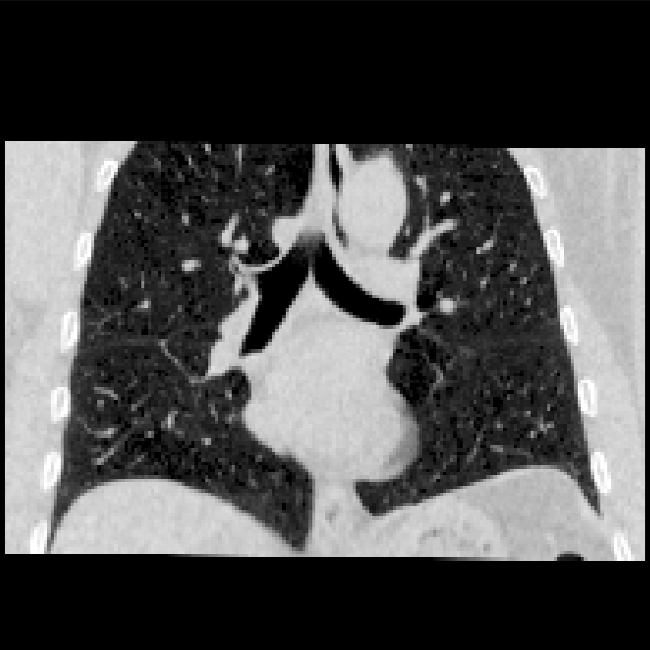} & \includegraphics[width=2cm]{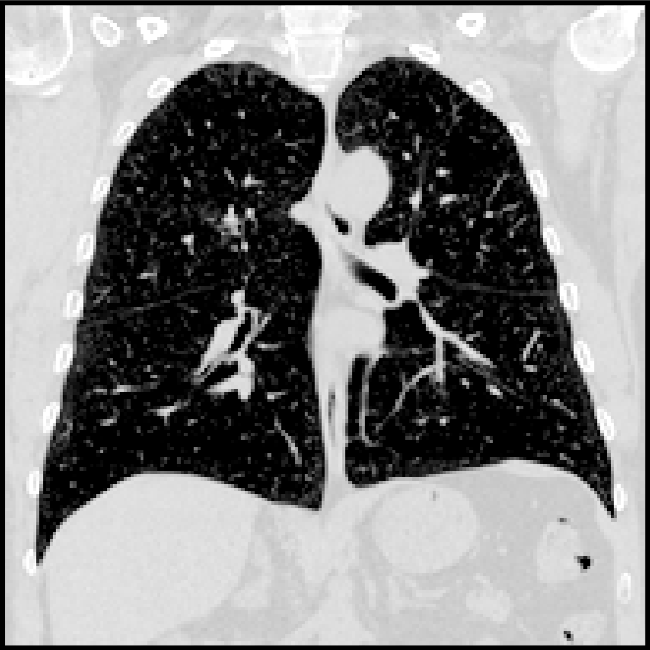}   \\
        \midrule
        Modalities & \multicolumn{2}{c}{MR T1w / CT} & \multicolumn{2}{c}{MR T1w / MR T1w} & \multicolumn{2}{c}{CT / CT} \\[0.1cm]
        Intra-/Inter-patient     & \multicolumn{2}{c}{Intra-patient}         & \multicolumn{2}{c}{Inter-patient} & \multicolumn{2}{c}{Intra-patient} \\[0.1cm]
        Resolution & \multicolumn{2}{c}{$192\stimes160\stimes192$} & \multicolumn{2}{c}{$160\stimes192\stimes224$}       & \multicolumn{2}{c}{$192\stimes192\stimes208$} \\[0.25cm]
        Voxel size & \multicolumn{2}{c}{$2\stimes2\stimes2\SI{}{\milli\meter}$} & \multicolumn{2}{c}{$1\stimes1\stimes1\SI{}{\milli\meter}$}       & \multicolumn{2}{c}{$1.75\stimes1.25\stimes1.75\SI{}{\milli\meter}$} \\[0.1cm]
        Cases (Train/Test) & \multicolumn{2}{c}{8/8} & \multicolumn{2}{c}{416/39} & \multicolumn{2}{c}{20/10} \\[0.1cm]
        Preprocessing & \multicolumn{2}{c}{canonical affine pre-align} & & & \multicolumn{2}{c}{affine pre-align} \\[0.05cm]
        & \multicolumn{2}{c}{crop/pad/resample} & & & \multicolumn{2}{c}{crop/pad/resample} \\[0.1cm]
        Annotations (Train/Test) & \multicolumn{2}{c}{4/9 anatomical labels} & \multicolumn{2}{c}{35/35 anatomical labels} & \multicolumn{2}{c}{--/100 landmarks/case} \\[0.1cm]
        Additional data & \multicolumn{2}{c}{90 unpaired MR/CT scans} & & &\multicolumn{2}{c}{lung masks} \\[0.05cm]
                        & \multicolumn{2}{c}{ROI masks} & & & & \\[0.1cm]
        Metrics & \multicolumn{2}{c}{DSC/DSC9/HD95} & \multicolumn{2}{c}{DSC/DSC30/HD95} & \multicolumn{2}{c}{TRE/TRE30} \\[0.05cm]
        & \multicolumn{2}{c}{SDlogJ/RT} & \multicolumn{2}{c}{SDlogJ/RT} & \multicolumn{2}{c}{SDlogJ/RT}\\[0.1cm]
        Challenges & \multicolumn{2}{l}{\textcolor{white}{$\sbullet$}\textcolor{OliveGreen}{$\sbullet$}\textcolor{RedOrange}{$\sbullet$}\textcolor{MidnightBlue}{$\sbullet$}\textcolor{Thistle}{$\sbullet$}\textcolor{Goldenrod}{$\sbullet$}\textcolor{Turquoise}{$\sbullet$}} & \multicolumn{2}{l}{\textcolor{white}{$\sbullet$}\textcolor{Maroon}{$\sbullet$}} & \multicolumn{2}{l}{\textcolor{white}{$\sbullet$}\textcolor{MidnightBlue}{$\sbullet$}\textcolor{Thistle}{$\sbullet$}\textcolor{Goldenrod}{$\sbullet$}\textcolor{black}{$\sbullet$}\textcolor{Turquoise}{$\sbullet$}} \\[0.1cm]
        \bottomrule[1.5pt]
    \end{tabular}
    \caption{Overview of all six Learn2Reg tasks addressing the imminent challenges of medical image registration: multi-modal scans~\textcolor{OliveGreen}{$\sbullet[2]$} (tasks with at least two different image modalities), few/noisy annotations~\textcolor{RedOrange}{$\sbullet[2]$} (less than five annotated anatomical structures for training cases), partial visibility~\textcolor{MidnightBlue}{$\sbullet[2]$} (restricted or cropped field of view for at least one image of a registration pair), small datasets~\textcolor{Thistle}{$\sbullet[2]$} (less than 30 training cases), large deformations~\textcolor{Goldenrod}{$\sbullet[2]$} (tasks with initial displacements of at least $10\SI{}{\centi\meter}$), small structures~\textcolor{Maroon}{$\sbullet[2]$} (tasks containing cases with target structures comprising less than 100 voxels), unsupervised registration~\textcolor{black}{$\sbullet[2]$} (no annotations for training cases) and missing correspondences~\textcolor{Turquoise}{$\sbullet[2]$} (e.g. due to removed organs, different field of views etc.)}.
    \label{tab:tasks}
\end{table*}

Learn2Reg consists of six clinically relevant complementary tasks (datasets). Table \ref{tab:tasks} summarises the dataset details, and we discuss them in detail below.

\paragraph*{\textbf{CuRIOUS}}
EASY-RESECT~\cite{Xiao_2020} is a simplified sub-set of the original RESECT dataset~\cite{xiao2017re}, previously used in the MICCAI CuRIOUS challenges~\cite{xiao2019curious2018}. The dataset contains 22 training and 10 testing subjects with low-grade brain gliomas, intended to help to develop MR vs. US registration algorithms to correct tissue shift in brain tumour resection. For the Learn2Reg challenge, we included T1w and T2-FLAIR MR scans, and spatially tracked intra-operative ultrasound volumes. All scans were acquired for routine clinical care of brain tumor resection procedures at St Olavs University Hospital (Trondheim, NO). Matching anatomical landmarks were annotated between T2-FLAIR MR and 3D ultrasound volumes~\cite{xiao2017re} to enable evaluation of the registration accuracy. During pre-processing, for each subject, the T1w scan is rigidly registered to the T2-FLAIR scan, and both scans are resampled to the same coordinate space as the 3D ultrasound volume yielding fixed voxel dimensions for all scans ($256\stimes256\stimes288$) at an isotropic resolution of approximately 0.5~mm. The registration to be carried out for this task was difficult for following reasons. First of all, it is a multimodal registration between MR and US images and the US images are typically noisier than the MR images. Furthermore, the pre-operative MR scans show a larger region of the brain whereas the intra-operative US volume was obtained to cover the entire tumor region after craniotomy but before dura opening.

\paragraph*{\textbf{Hippocampus MR}}
This dataset consists of 394 MR scans of the hippocampus region acquired in 90 healthy adults and 105 adults with non-affective psychotic disorder taken from the Psychiatric Genotype/Phenotype Project data repository at Vanderbilt University Medical Center (VUMC). The hippocampus head and tail were manually traced in all scans. The ability to establish correspondences for small structures between patients is particularly important for accurate population analysis. Previous to the Learn2Reg challenge, the dataset was used as part of the Medical Segmentation Decathlon~\cite{antonelli2021medical}. Due to its small volumetric size and large training dataset with two anatomical labels, Hippocampus MR appeared to be a good entry-level task for learning-based registration approaches.

\paragraph*{\textbf{Abdomen CT-CT}}\label{sec:abdomenctct}
This task tackles inter-patient registration of abdominal CT scans, which enables statistical modelling of variations of organs for abnormality detection, and can provide a canonical atlas space for further investigations. The dataset contains 50 abdominal CT scans (30/20 for training/testing) with 13 manually labelled anatomical structures: spleen, right/left kidney, gall bladder, esophagus, liver, stomach, aorta, inferior vena cava, portal and splenic vein, pancreas and left/right adrenal gland. Data acquisition and annotation protocols are detailed in~\cite{xu2016evaluation}. The images were registered affinely in  a groupwise manner and resampled to the same voxel resolution and spatial dimensions ($192\stimes160\stimes256$).

\paragraph*{\textbf{Abdomen MR-CT}}
The data was compiled from public studies of the cancer imaging archive (TCIA)~\cite{clark2013cancer} that contained paired scans of both MR and CT from the same patients. In particular, 16 MR and CT scans from the following studies, TCGA-KIRC~\cite{akin2016radiology}, TCGA-KIRP~\cite{linehan2016radiology}, and TCGA-LIHC~\cite{erickson2016radiology}, are included in Learn2Reg - that cover routine diagnostic scans and follow-up imaging for kidney surgery. The data has been resampled to an isotropic resolution of 2mm, and cropped and padded to achieve voxel dimensions of 192x160x192. We have also manually traced 3D segmentation masks for the liver, spleen, left and right kidney. All scans were pre-aligned using a groupwise affine registration based on the deeds-linear algorithm~\cite{heinrich2012deeds}. Additional unpaired and segmented training data from two further challenges - BCV-CT~\cite{xu2016evaluation} and CHAOS-MR~\cite{CHAOSdata2019,kavur2021chaos} - were provided for pre-training.

\paragraph*{\textbf{OASIS}}
The task employed 416 3D whole-brain MR scans from the Open access series of imaging studies (OASIS)~\cite{marcus2007open}, a cross-sectional MR data study with a wide range of participants from young, middle-aged, nondemented, and demented older adults. The clinical relevance of this inter-patient registration task lies in quantitative brain analysis, which is of utmost importance for a better understanding of the human brain and for the analysis of various brain diseases. Standard brain MR pre-processing including skull-stripping \revised{(optional)}, normalisation, pre-alignment, and resampling was performed. Semi-automatic labels with manual corrections of 35 cortical and subcortical brain structures were generated using FreeSurfer~\cite{fischl2012freesurfer}. For details on data curation, see~\cite{hoopes2021hypermorph}.

\paragraph*{\textbf{Lung CT}}
The aim of the lung CT task was the registration of expiration to inspiration CT scans of the lung. Establishing correspondences between longitudinal lung scans can help to monitor disease progression, estimate motion in radiotherapy planning or enable direct assessment of lung ventilation. The data consists of 20 training~\cite{hering_alessa_2020_3835682} and 10 test scan pairs~\cite{hering_alessa_2020_4048761}. The scans were acquired at the Dept. of Radiology at the Radboud University Medical Center, Nijmegen, NL. All pairs were affinely pre-registered and resampled to an image size of $192\stimes192\stimes208$. Lung segmentation masks and keypoints were provided as additional training information. The complexity of this registration task is manifold. First, the fields of view of the fixed and moving scan differ largely since the lungs in the expiration scan are not fully visible. Second, the scale of the motion within the lungs can often be larger than the anatomical structures (vessels and airways) themselves. Therefore, a registration method needs to estimate large displacements that account for substantial breathing motion and also align small structures like individual pulmonary blood vessels precisely. To measure the accuracy manual landmarks are used that are typically located at the boundary or bifurcation of vessels, airways, and parenchyma.

\subsection{Challenge Design}
To provide a comprehensive evaluation of the registration performance, we consider a number of complementary metrics (see section \ref{sec:metrics}) that assess the accuracy, robustness, plausibility, and speed of the algorithms. For final task ranks, we further consider the significance of differences in results. The detailed ranking scheme is described in section~\ref{sec:ranking}.

\subsubsection{Metrics}
\label{sec:metrics}
\paragraph*{\textbf{DSC}} The Dice similarity coefficient (DSC) measures the overlap of two sets of segmentation labels (on the fixed and warped moving scan).

\paragraph*{\textbf{DSC30}} To assess robustness, the DSC30 metric considers the 30th percentile in DSC scores over all anatomical structures and cases.

\paragraph*{\textbf{DSC9}} DSC9 is a special metric introduced for the Abdomen MR-CT task, to asses the effect of label bias. It is evaluated on 9 additional anatomical labels, that were not available during training.

\paragraph*{\textbf{HD95}} The Hausdorff distance (HD) indicates the maximum distance in a metric space (here: Euclidean space, distance specified in millimetres (mm)) between two sets of surfaces (segmentation labels on the fixed and warped moving scan). For a robust score, we consider the 95th percentile instead of the maximum distance (HD95).

\paragraph*{\textbf{TRE}} The target registration error (TRE) is defined as the euclidean distance (in millimetres (mm)) between corresponding landmarks in the warped fixed and moving scan.

\paragraph*{\textbf{TRE30}} Similar to the DSC30 score the TRE30 metric collects the 30th percentile of largest landmark distances.

\paragraph*{\textbf{SDlogJ}} The plausibility (smoothness) of a displacement field is captured using the standard deviation of the logarithm of the Jacobian determinant (SDlogJ) of the displacement field \cite{leow2007statistical,kabus2009evaluation}. The Jacobian is calculated by a central differencing approximation.

\paragraph*{\textbf{RT}} In addition, we are able to measure the test-time registration runtime (RT) on the same hardware (CPU: Xeon Silver 4210R, GPU: Quadro RTX 8000), when methods are submitted as a Docker container. Start and stop times are the loading of the first scan and writing of the displacement field to disk, respectively. 

\subsubsection{Ranking Scheme}
The ranking scheme is based on the ranking scheme of the Medical Decathlon\footnote{\url{http://medicaldecathlon.com}}. We rank methods using statistically significantly different results. For each metric applied in a task, methods are compared against each other (Wilcoxon signed rank test with p\textless0.05), ranked based on the number of "won" comparisons and finally mapped to a numerical metric rank score between 0.1 and 1 (with possible score sharing). A task rank score is then obtained as the geometric mean of individual metric rank scores. All methods for which no metric is available (not submitted to the task, no Docker container submitted) share the lowest possible metric rank score of 0.1.

\label{sec:ranking}

\section{Challenge Entries}\label{sec:entries}
\revised{In 2020, ten teams submitted their solutions. The total number of teams increased to 21 in 2021. Counting the submissions task-wise results in 65 unique challenge entries.}
\removed{
In phase 1, performed using the grand-challenge.org platform, 17 teams submitted displacement fields in 2020 and 22 teams in 2021.
In phase 2, two teams submitted displacement fields in 2020 and eight teams submitted their algorithms as Docker images. In 2021, three teams submitted displacement fields and 12 teams submitted a Docker container. Only algorithms that participated in both phases of at least one year were included in this paper.} Table~\ref{tab:overviewParticipants} provides a summary of important information. Below is a brief description of each of the 21 submissions. For more details, please refer to the respective articles in the proceedings of the MICCAI Learn2Reg workshops.
\begin{table*}[!h]
    \setlength{\tabcolsep}{2pt}
    \renewcommand{\arraystretch}{0.75}
    \centering
    \begin{tabular}{r?c|c|c|c|c|c?c|c?c|c|c|c|c|c|c?c|c?c|c|c?c?l}
        \toprule[1.5pt]
        & \multicolumn{6}{c?}{Tasks} & \multicolumn{2}{c?}{Type} & \multicolumn{7}{c?}{Objectives} & \multicolumn{2}{c?}{Reg.} & \multicolumn{3}{c?}{Optimis.} && Misc. (architectures, add. objectives, etc.) \\
        & \rotatebox{90}{CuRIOUS} & \rotatebox{90}{Hippocampus MR} & \rotatebox{90}{Abdomen CT-CT} & \rotatebox{90}{Abdomen MR-CT} & \rotatebox{90}{OASIS} & \rotatebox{90}{Lung CT} & \rotatebox{90}{Conventional} & \rotatebox{90}{Deep Learning} & \rotatebox{90}{NCC} & \rotatebox{90}{MIND-SSC} & \rotatebox{90}{NGF} & \rotatebox{90}{MI} &\rotatebox{90}{Dice} & \rotatebox{90}{Keypoints} & \rotatebox{90}{Consistency (Inv./Cycl.)} & \rotatebox{90}{Diffusion} & \rotatebox{90}{Curvature} & \rotatebox{90}{Adam} & \rotatebox{90}{Convex} & \rotatebox{90}{L-BFGS/Gauss-Newton} &\rotatebox{90}{Open source} &\\
        \midrule                                                                                            %
        3Idiots \textcolor{3Idiots}{\scalebox{1.25}{$\blacksquare$}}        &   &   &   &   &\x &   &   &\x &   &   &   &\x &\x &   &   &\x &   &\x &   &   &&Voxelmorph; SSD;\\
        Bailiang \textcolor{Bailiang}{\scalebox{1.25}{$\blacksquare$}}      &   &   &   &   &\x &   &   &\x &\x &\x &   &   &\x &   &   &\x &   &\x &   &   &\href{https://github.com/BailiangJ/learn2reg2021_task3}{\faicon{github}}&DeepRegNet;\\
        ConvexAdam \textcolor{ConvexAdam}{\scalebox{1.25}{$\blacksquare$}}  &\y &\x &\x &\y &\x &\y &\z &\z &   &\z &   &   &\z &   &\z &\x &   &\z &\z &   &\href{https://github.com/multimodallearning/ConvexAdam}{\faicon{github}}&UNet; Dense corr.;\\
        corrField$^*$ \textcolor{corrField}{\scalebox{1.25}{$\blacksquare$}}    &\y &\y &\y &\y &\y &\y &\x &   &   &\x &   &   &   &   &\x &\x &   &   &   &   &\href{https://grand-challenge.org/algorithms/corrfield/
}{\faicon{github}}&Dense corr.;\\
        Driver \textcolor{Driver}{\scalebox{1.25}{$\blacksquare$}}          &   &   &   &\x &\x &\x &   &\x &\x &   &   &\x &\z &\z &   &   &   &\x &   &   &&PCNet; Cross-entropy loss;\\
        Epicure \textcolor{Fourcade}{\scalebox{1.25}{$\blacksquare$}}       &   &   &   &   &   &\y &\x &   &\x &   &   &   &   &   &   &   &   &   &   &   &&Bending energy regularisation;\\
        Estienne \textcolor{Estienne}{\scalebox{1.25}{$\blacksquare$}}      &   &\x &\x &   &   &   &   &\x &\x &   &   &   &\x &   &   &\x &   &\x &   &   &\href{https://github.com/TheoEst/abdominal_registration
}{\faicon{github}}&UNet; Multi-Task learning; \\
        Gunnarsson \textcolor{Gunnarsson}{\scalebox{1.25}{$\blacksquare$}}  &\y &\x &\x &   &   &\y &   &\x &\x &   &   &   &\x &   &   &\x &   &\x &   &   &\href{https://github.com/ngunnar/learning-a-deformable-registration-pyramid}{\faicon{github}}&PWC;\\
        Imperial \textcolor{Imperial}{\scalebox{1.25}{$\blacksquare$}}      &   &   &   &\x &\x &\y &   &\x &   &   &\x &   &\x &\x &   &\x &   &\x &   &   &\href{https://github.com/biomedia-mira/istn
}{\faicon{github}}&Structure-guided loss;\\
        Joutard \textcolor{Joutard}{\scalebox{1.25}{$\blacksquare$}}        &   &   &\x &   &   &   &   &\x &   &   &   &   &\x &   &   &\x &   &\x &   &   &&UNet; EDT similarity; Dense corr.;\\
        LapIRN \textcolor{LapIRN}{\scalebox{1.25}{$\blacksquare$}}          &\y &\x &\x &\x &\x &\y &\z &\z &\x &\z &   &\z &\z &   &   &\x &   &\x &   &   &\href{https://github.com/cwmok/Conditional_LapIRN}{\faicon{github}}&UNet; Conditional NN;\\
        LaTIM \textcolor{Jaouen}{\scalebox{1.25}{$\blacksquare$}}           &   &   &   &\y &\y &\y &\x &   &   &   &   &   &   &   &   &   &   &   &   &   &&Directional representations;\\
        Lifshitz \textcolor{Lifshitz}{\scalebox{1.25}{$\blacksquare$}}      &   &   &   &   &   &\y &   &\x &\x &   &   &   &   &   &\x &   &   &   &   &   &&Unrolled $L_1$ regulariser; Dense corr.;\\
        lWM \textcolor{IWM}{\scalebox{1.25}{$\blacksquare$}}                &   &\x &   &   &\x &   &   &\x &   &\x &   &   &\x &   &\x &\x &   &\x &   &   &&2-stream NN;\\
        MEVIS \textcolor{MEVIS}{\scalebox{1.25}{$\blacksquare$}}            &\y &\x &\y &\y &\y &\y &\z &\z &   &   &\x &   &\z &\z &   &   &\x &\z &   &\z &&\\
        Multi-brain \textcolor{Brudfors}{\scalebox{1.25}{$\blacksquare$}}   &   &   &   &\y &\y &\y &\x &   &   &   &   &   &   &   &   &\x &   &   &   &\x   &\href{https://github.com/WTCN-computational-anatomy-group/mb
}{\faicon{github}}&Groupwise; Bayesian modelling;\\
        NiftyReg$^*$ \textcolor{NiftyReg}{\scalebox{1.25}{$\blacksquare$}}      &\y &\y &\y &\y &\y &\y &\x &   &\z &\z &   &   &   &   &   &   &   &   &   &   &\href{https://github.com/KCL-BMEIS/niftyreg
}{\faicon{github}}&Bending and Jacobian regularisation;\\
        PDD-Net$^*$ \textcolor{PDD-Net}{\scalebox{1.25}{$\blacksquare$}}        &\y &\y &\x &   &   &\y &   &\x &   &\x &   &   &\z &   &   &\x &   &\x &   &   &\href{https://github.com/multimodallearning/pdd_net}{\faicon{github}}&Dense Corr.;\\
        PIMed \textcolor{PIMed}{\scalebox{1.25}{$\blacksquare$}}            &   &   &\x &\x &\x &\y &\z &\z &\x &   &   &   &\z &   &   &   &   &\z &   &\z &&UNet; SSTVD similarity; Dense corr.;\\
        Thorley \textcolor{Thorley}{\scalebox{1.25}{$\blacksquare$}}        &   &   &   &   &\x &   &\x & \x  &   &   &   &   & \x  &   &   & \x  &   &   & \x  &   & & UNet; SAD\\
        Winter \textcolor{Winter}{\scalebox{1.25}{$\blacksquare$}}          &   &   &   &\y &\y &\y &\z &\z &\x &\x &   &\x &   &   &   &\x &   &\x &   &   &\href{https://github.com/WinterPan2017/ADLReg}{\faicon{github}}&Voxelmorph;\\
        \bottomrule[1.5pt]
    \end{tabular}
    \caption{Methodological overview of all Learn2Reg methods. An entry in the table indicates agreement with the corresponding heading. Unsupervised and supervised challenge entries are marked with a \y{} and \x{} symbol in the \textit{Tasks} subgroup. If a challenge entry uses different approaches for different tasks or mixes them within the method (e.g. Deep Learning + Instance Optimisation) we marked the property with a \z{} symbol. All baseline methods are marked with an *. For detailed descriptions of the methods see Section \ref{sec:entries} and the associated references.
    }
    \label{tab:overviewParticipants}
\end{table*}

\paragraph*{\textbf{3Idiots \textcolor{3Idiots}{\scalebox{1.25}{$\blacksquare$}}}}
\cite{han2021deformable} employs deep-learning-based approach using a hybrid similarity loss consisting of intensity (SSD), statistical (MI), and label-based (Dice+L1) penalties. A Voxelmorph~\cite{balakrishnan2019voxelmorph} model with an increased number of feature channels and halved output resolution is trained in a patch-wise manner and applied to the OASIS task.

\paragraph*{\textbf{Bailiang \textcolor{Bailiang}{\scalebox{1.25}{$\blacksquare$}}}}
\cite{DeepRegNet} addresses OASIS and is based on the DeepRegNet framework from Project-MONAI. The input of the encoder is the concatenation of fixed and moving images. A dense vector field (DVF) is predicted from summing over different level decoders and integrated using scaling and squaring. The loss function is composed of LNCC, MIND-SCC, Dice, and a diffusion regulariser. \url{https://github.com/BailiangJ/learn2reg2021_task3}

\paragraph*{\textbf{ConvexAdam \textcolor{ConvexAdam}{\scalebox{1.25}{$\blacksquare$}}}}
\cite{siebert2021fast} proposes a decoupling of deep learning for semantic feature extraction and the conventional optimisation. They combine a single-level dense discretised displacement correlation with large capture range and convex global optimisation with a local gradient-based instance refinement using the Adam optimiser. The method is applied to all six tasks and uses diffusion regularisation, an inverse-consistency constraint, and MIND similarity. The method extends the input features to learned label-supervised representations for inter-patient tasks: Abdomen CT-CT, Hippocampus MR, and OASIS.
\url{https://github.com/multimodallearning/ConvexAdam}

\paragraph*{\textbf{corrField \textcolor{corrField}{\scalebox{1.25}{$\blacksquare$}}}}
A faster implementation (from~\cite{hansen2021graphregnet}) of the corrField method~\cite{heinrich2015estimating} is introduced as a non-learning based unsupervised baseline. The method estimates sparse correspondences on image-based Förstner keypoints with exact message passing on a minimum spanning tree. MIND-SSC features are used for the similarity term.
\url{https://grand-challenge.org/algorithms/corrfield/}

\paragraph*{\textbf{Driver \textcolor{Driver}{\scalebox{1.25}{$\blacksquare$}}}}
\cite{lv2021joint} uses a dual-encoder UNet backbone with separated multi-scale feature extractors that comprises Deformation Field Integration (DFI) and non-rigid feature fusion (NFF) modules. It produces multi-scale sub-fields that progressively align fixed and moving features. The overall framework comprises a rigid transform network and MI or LNCC similarity, weak label-supervision and regularisation.

\paragraph*{\textbf{Epicure \textcolor{Fourcade}{\scalebox{1.25}{$\blacksquare$}}}}
\cite{fourcade2020using} addresses the lung CT task using an iterative registration approach based on the Elastix toolbox \cite{klein2009elastix} optimising the object function that is composed of the NCC similarity and a bending energy penalty term.

\paragraph*{\textbf{Estienne \textcolor{Estienne}{\scalebox{1.25}{$\blacksquare$}}}}
\cite{estienne2020journal,estienne2020learn2reg} combines a diffeomorphic symmetric spatial transformer network with a embedding merging step, that eases the learning by subtracting the embeddings of separately encoded fixed and moving scans and thereby leveraging the prior knowledge that swapped inputs should yield negated velocity fields. They extend the label-based pre-training by including additional public datasets with at least partial overlap in segmentation classes, using segmentation masks produced by a CNN. \url{https://github.com/TheoEst/abdominal_registration}

\paragraph*{\textbf{Gunnarsson \textcolor{Gunnarsson}{\scalebox{1.25}{$\blacksquare$}}}}
\cite{gunnarsson2020learning} proposes an end-to-end learning-based 3D registration method inspired by the PWC-Net~\cite{sun2018pwc}. The method estimates and refines a displacement field from a cost volume at each level of a CNN downsampling pyramid and is supervised by a similarity (NCC) and/or segmentation (DICE) loss, as well as a smoothness penalty. The network is trained and evaluated on scan pairs from the four tasks of the 2020 challenge (CuRIOUS, Lung CT, Abdomen CT-CT and Hippocampus MR). \url{https://github.com/ngunnar/learning-a-deformable-registration-pyramid}

\paragraph*{\textbf{Imperial \textcolor{Imperial}{\scalebox{1.25}{$\blacksquare$}}}}
\cite{lee2019istn} uses Image-and-Spatial Transformer Networks (ISTN) as the backbone of their method. In the ISTN, the fixed and moving images are first separately processed by the ITN to generate a segmentation mask and a feature map of the input image. Subsequently, both feature maps are used by the STN to predict the displacement field. The loss function consists of a structural-guided and image similarity and a regularisation loss. \url{https://github.com/biomedia-mira/istn}

\paragraph*{\textbf{Joutard \textcolor{Joutard}{\scalebox{1.25}{$\blacksquare$}}}}
Joutard addresses the Abdomen CT-CT task with a weakly supervised deep learning approach. A CNN extracts features from the fixed and moving image, which are concatenated with their spatial image coordinates. The feature distributions for each spatial location are then matched between the two images which yield a correspondence matrix from which the average displacement can be derived. The network is supervised by a segmentation (Dice) and a regularisation (L2 norm on gradients) loss.

\paragraph*{\textbf{LapIRN \textcolor{LapIRN}{\scalebox{1.25}{$\blacksquare$}}}}
\cite{mok2020large,mok2021conditional} propose an image registration method based on Laplacian pyramid registration networks to overcome the large inter-and intra-variations of anatomical structures in the input scans. For the 2021 tasks (Abdomen MR-CT, OASIS and Lung CT),~\cite{mok2021conditional} extended their initial approach \cite{mok2020large} by adding a conditional module that enables the input of the regularisation hyperparameter so that the different solutions for different hyperparameter values can be captured by a single convolutional neural network.\url{https://github.com/cwmok/Conditional_LapIRN}

\paragraph*{\textbf{LaTIM \textcolor{Jaouen}{\scalebox{1.25}{$\blacksquare$}}}}
\cite{Jaouen2021Reg} addresses the Abdomen CT-CT tasks using an iterative technique exploiting vector-valued directional image representations. The method is implemented within the Elastix framework.

\paragraph*{\textbf{Lifshitz \textcolor{Lifshitz}{\scalebox{1.25}{$\blacksquare$}}}}
\cite{lifshitz2021cost} proposes a deep-learning-based solution for the Lung CT task that comprises a 3D extension of ARFlow~\cite{liu2020ARFlow} with multi-resolution warping, displacement correlation, and flow estimation. To address edge-preservation of sliding motion an unrolling of the total variation (L1) regularisation loss using variable substitution is proposed.

\paragraph*{\textbf{lWM \textcolor{IWM}{\scalebox{1.25}{$\blacksquare$}}}}
lWM employs a deep-learning-based registration method for the Hippocampus MR and the OASIS task. For the Hippocampus MR task, they use sequential deformation field composition, while the solution for the OASIS task uses an image pyramid separately applied to both input images and a UNet with residual blocks. The objective function includes MIND, Dice, inverse consistency and diffusion losses.

\paragraph*{\textbf{MEVIS \textcolor{MEVIS}{\scalebox{1.25}{$\blacksquare$}}}}
The submission of MEVIS~\cite{hager2020variable,hering2021fraunhofer} solves all tasks besides the Hippocampus MR task using a conventional method and build on cost functions and losses made up from several terms that are selected for the specific task. The method use a coarse-to-fine multi-level iterative registration scheme where a Gaussian image pyramid is generated for both images to obtain downsampled and smoothed images. On each level, a  quasi-Newton L-BFGS optimisation is used. For the Hippocampus task, a deep learning approach with a weakly supervised trained UNet is applied using the same cost function as in the conventional approach.

\paragraph*{\textbf{Multi-brain \textcolor{Brudfors}{\scalebox{1.25}{$\blacksquare$}}}}
\cite{brudfors2020flexible} uses groupwise, fully unsupervised registration techniques based on Bayesian modelling and Gauss-Newton optimisation, which learns priors over image intensities and spatial tissue classes. The method requires no pre-processing of the imaging data and does not utilise label information. The method is applied to Abdomen MR-CT, OASIS, and Lung CT. \url{https://github.com/WTCN-computational-anatomy-group/mb}

\paragraph*{\textbf{NiftyReg \textcolor{NiftyReg}{\scalebox{1.25}{$\blacksquare$}}}}\cite{modat2010fast} is applied as conventional baseline for all tasks without label supervision using NCC for CuRIOUS and otherwise MIND as similarity metric. Both bending and Jacobian regularisation penalties are applied and the number of pyramid levels is restricted to yield competitive run times. \url{https://github.com/KCL-BMEIS/niftyreg}

\paragraph*{\textbf{PDD-Net \textcolor{PDD-Net}{\scalebox{1.25}{$\blacksquare$}}}}
The PDD-Net~\cite{heinrich2019closing,heinrich2020highly} is used as a baseline method. It uses a deformable convolutional network to extract image features and compute a six-dimensional dissimilarity tensor (three spatial + three displacement dimensions). A smooth displacement field is obtained from the dissimilarities by mean field inference over spatial dimensions and approximated min-convolutions over displacement dimensions. The method is adapted to four challenge tasks (CuRIOUS, Hippocampus MR, Abdomen CT-CT, and Lung CT). \url{https://github.com/multimodallearning/pdd_net}

\paragraph*{\textbf{PIMed \textcolor{PIMed}{\scalebox{1.25}{$\blacksquare$}}}}
PIMed uses a multi-slice segmentation network that yields anatomical maps and is employed for Abdomen MR-CT and Abdomen CT-CT in conjunction with a NCC loss and optimised using 1) a translation only and 2) a diffeomorphic deformation model. They adapt a residual VoxelMorph model with weak supervision for OASIS. For lung CT, they apply a conventional method with geodesic density regression and adaptation of intensities to lung tissue density~\cite{shao2021geodesic}.

\paragraph*{\textbf{Thorley}  \textcolor{Thorley}{\scalebox{1.25}{$\blacksquare$}} }
\revised{The submission from the University of Birmingham (UoB) team tackled the OASIS task using an iterative coarse-to-fine registration scheme, optimizing the classical SAD difference term and a third-order diffusion displacement regularizer. Additionally, they decomposed the transformation into the composition of a series of small non-stationary velocity fields, and solved the convex optimization using the Nesterov accelerated ADMM~\cite{thorley2021nesterov} with closed-form solutions. An additional post processing step using a UNet supervised with dice and diffusion loss was used to further refine the displacement fields produced by the iterative optimization.}

\paragraph*{\textbf{Winter \textcolor{Winter}{\scalebox{1.25}{$\blacksquare$}}}}
Winter addresses the Abdomen MR-CT, OASIS and Lung CT task by employing a conventional method for Lung CT and a attention-based deep-learning-based registration method for Abdomen MR-CT and OASIS brain. For the Abdomen MR-CT task, a two-step approach is applied that first aligns the provided ROI masks. \url{https://github.com/WinterPan2017/ADLReg}

\section{Additional Experiments}
\paragraph*{\textbf{Label Bias}}
Previous publications on learning-based registration have already discussed the possibility of bias towards anatomies that are used both for training and evaluation~\cite{balakrishnan2019voxelmorph}. While this bias is intrinsic to all segmentation approaches, registration is often used as a more generalistic tool in clinical applications that may require accurate alignment of structures that are not defined a priori. To study the effect of adding anatomical labels to the evaluation that were not present during method development and training, we extended both abdomen tasks. For the inter-patient CT-CT registration we included the duodenum with the manual annotations provided by~\cite{gibson2018automatic}, for the intra-patient MR-CT task we extended the predominantly large organs by five smaller ones: gallbladder, stomach, aorta, portal vein, pancreas (semi-automatically generated using a nnUNet trained on the VISCERAL gold corpus \cite{jimenez2016cloud}).

\paragraph*{\textbf{Unsupervised Registration}}
The top-performing methods are all modular in their use of segmentation labels for supervision. As analysed in the label bias experiment, there is a risk of over-fitting registration performance to the chosen subset of manually annotated anatomies. We, hence, compared the unsupervised counterparts of the following methods: LapIRN and ConvexAdam. ConvexAdam already uses an unsupervised method for all three intra-patient tasks, and LapIRN for CuRIOUS and Lung CT. Therefore the additional comparisons are restricted to the abdomen and brain.

\paragraph*{\textbf{Transferability}}
A robust registration method should work well for all scan pairs regardless of acquisition parameters and thus on comparable datasets. A limitation of deep-learning-based methods might be that they reach higher accuracy on the dataset they are trained on and show a considerable loss of accuracy on other data. As in~\cite{hering2021cnn,hoffmann2021synthmorph}, we evaluate the transferability of methods submitted to the lung CT-CT task by registering the DIRLab 4DCT~\cite{castillo2009framework} scan pairs. The scans are preprocessed in the same way as the scans of the lung CT-CT task. The evaluation is based on the target registration error of the landmarks and the smoothness of the deformation field. Furthermore, this experiment allows comparison to a variety of other lung registration methods, as the DIRLAB data set is often used as a benchmark (please note that the reduced resolution leads to a general deterioration of TRE of $\sapprox$0.3mm).

\section{Results}
\begin{table*}
    \RawFloats
    \small
    \setlength{\tabcolsep}{0.000001\textwidth}
    \setlength{\fboxsep}{2pt}
  \floatsetup{floatrowsep=qquad, captionskip=4pt}
  \begin{floatrow}[2]
        \ttabbox
       {\begin{tabularx}{0.49\textwidth}{rCCCCc}
        \toprule[1.5pt]
        & TRE$\downarrow$ & TRE30$\downarrow$ & SDLogJ$\downarrow$ & RT$\downarrow$ & Rank$\uparrow$ \\
        \midrule
        Initial & 6.38 & 12.00 & & & \\
        \midrule
        corrField \textcolor{corrField}{\scalebox{1.25}{$\blacksquare$}} & 2.84 & 5.29 & 0.00 & 2.70 & \colorbox{gold}{0.85$\pm$0.03}\\
        PDD-Net \textcolor{PDD-Net}{\scalebox{1.25}{$\blacksquare$}} & 3.08 & 6.28 & 0.00 & 8.21 & \colorbox{silver}{0.83$\pm$0.03}\\
        ConvexAdam \textcolor{ConvexAdam}{\scalebox{1.25}{$\blacksquare$}} & 3.31 & 5.82 & 0.00 & 1.33 & \colorbox{bronze}{0.77$\pm$0.04}\\
        NiftyReg \textcolor{NiftyReg}{\scalebox{1.25}{$\blacksquare$}} & 4.09 & 7.85 & 0.00 & 23.1 & 0.56$\pm$0.06\\
        LapIRN \textcolor{LapIRN}{\scalebox{1.25}{$\blacksquare$}} & 5.67 & 11.1 & 0.00 & 34.8 & 0.49$\pm$0.06\\
        MEVIS \textcolor{MEVIS}{\scalebox{1.25}{$\blacksquare$}} & 6.55 & 10.4 & 0.00 & 57.8 & 0.42$\pm$0.04\\
        Gunnarsson \textcolor{Gunnarsson}{\scalebox{1.25}{$\blacksquare$}} & 7.1 & 10.1 & 0.14 & 42.2 & 0.19$\pm$0.01\\
        \textcolor{White}{\scalebox{1.25}{$\blacksquare$}}&&&&&\\
        \textcolor{White}{\scalebox{1.25}{$\blacksquare$}}&&&&&\\
        \bottomrule[1.5pt]
      \end{tabularx}}
    {\caption[CuRIOUS]{CuRIOUS}
      \label{tabCuRIOUS}}
    \hfill%
    \ttabbox%
    {\begin{tabularx}{0.49\textwidth}{rCCCCCc}
        \toprule[1.5pt]
        & DSC$\uparrow$ & DSC30$\uparrow$ & HD95$\downarrow$ & SDLogJ$\downarrow$ & RT$\downarrow$ & Rank$\uparrow$ \\
        \midrule
        Initial & 0.55 & 0.36 & 3.91 & & & \\
        \midrule
        LapIRN \textcolor{LapIRN}{\scalebox{1.25}{$\blacksquare$}} & 0.88 & 0.86 & 1.30 & 0.05 & 1.03 & \colorbox{gold}{0.93$\pm$0.01}\\
        MEVIS \textcolor{MEVIS}{\scalebox{1.25}{$\blacksquare$}} & 0.85 & 0.84 & 1.55 & 0.05 & 0.59 & \colorbox{silver}{0.78$\pm$0.03}\\
        ConvexAdam \textcolor{ConvexAdam}{\scalebox{1.25}{$\blacksquare$}} & 0.84 & 0.83 & 1.85 & 0.07 & 0.48 & \colorbox{bronze}{0.75$\pm$0.04}\\
        lWM \textcolor{IWM}{\scalebox{1.25}{$\blacksquare$}} & 0.79 & 0.76 & 2.20 & 0.08 & 0.80 & 0.63$\pm$0.04\\
        Estienne \textcolor{Estienne}{\scalebox{1.25}{$\blacksquare$}} & 0.85 & 0.84 & 1.51 & 0.09 & 1.46 & 0.62$\pm$0.04\\
        PDD-Net \textcolor{PDD-Net}{\scalebox{1.25}{$\blacksquare$}} & 0.78 & 0.76 & 2.23 & 0.07 & 0.35 & 0.58$\pm$0.04\\
        NiftyReg \textcolor{NiftyReg}{\scalebox{1.25}{$\blacksquare$}} & 0.76 & 0.72 & 2.72 & 0.09 & 4.75 & 0.37$\pm$0.03\\
        corrField \textcolor{corrField}{\scalebox{1.25}{$\blacksquare$}} & 0.72 & 0.68 & 2.89 & 0.05 & 1.20 & 0.34$\pm$0.02\\
        Gunnarsson \textcolor{Gunnarsson}{\scalebox{1.25}{$\blacksquare$}} & 0.74 & 0.67 & 2.82 & 0.16 & 22.0 & 0.25$\pm$0.01\\
        \bottomrule[1.5pt]
      \end{tabularx}}
    {\caption[Hippocampus MR]{Hippocampus MR}
      \label{tabHippocampusMR}}
  \end{floatrow}
  \vspace*{1em}
  \begin{floatrow}[2]
    \ttabbox%
    {\begin{tabularx}{0.49\textwidth}{rCCCCCc}
        \toprule[1.5pt]
        & DSC$\uparrow$ & DSC30$\uparrow$ & HD95$\downarrow$ & SDLogJ$\downarrow$ & RT$\downarrow$ & Rank$\uparrow$ \\
        \midrule
        Initial & 0.28 & 0.04 & 21.78 & & & \\
        \midrule
        ConvexAdam \textcolor{ConvexAdam}{\scalebox{1.25}{$\blacksquare$}} & 0.69 & 0.45 & 11.03 & 0.06 & 2.75 & \colorbox{gold}{0.94$\pm$0.01}\\
        LapIRN \textcolor{LapIRN}{\scalebox{1.25}{$\blacksquare$}} & 0.67 & 0.47 & 12.51 & 0.12 & 3.80 & \colorbox{silver}{0.82$\pm$0.03}\\
        Estienne \textcolor{Estienne}{\scalebox{1.25}{$\blacksquare$}} & 0.69 & 0.51 & 11.77 & 0.18 & 8.23 & \colorbox{bronze}{0.67$\pm$0.08}\\
        MEVIS \textcolor{MEVIS}{\scalebox{1.25}{$\blacksquare$}} & 0.51 & 0.24 & 18.21 & 0.14 & 3.49 & 0.60$\pm$0.04\\
        corrField \textcolor{corrField}{\scalebox{1.25}{$\blacksquare$}} & 0.49 & 0.24 & 17.22 & 0.28 & 5.40 & 0.53$\pm$0.04\\
        PIMed \textcolor{PIMed}{\scalebox{1.25}{$\blacksquare$}} & 0.49 & 0.23 & 15.75 & 0.05 & & 0.49$\pm$0.04\\
        PDD-Net \textcolor{PDD-Net}{\scalebox{1.25}{$\blacksquare$}} & 0.49 & 0.24 & 17.75 & 0.41 & 6.06 & 0.44$\pm$0.02\\
        Joutard \textcolor{Joutard}{\scalebox{1.25}{$\blacksquare$}} & 0.40 & 0.13 & 17.25 & 0.05 & 3.67 & 0.42$\pm$0.01\\
        NiftyReg \textcolor{NiftyReg}{\scalebox{1.25}{$\blacksquare$}} & 0.45 & 0.20 & 20.70 & 0.36 & 17.1 & 0.36$\pm$0.02\\
        Gunnarsson \textcolor{Gunnarsson}{\scalebox{1.25}{$\blacksquare$}} & 0.43 & 0.17 & 18.55 & 0.13 & 31.5 & 0.33$\pm$0.02\\
        \textcolor{White}{\scalebox{1.25}{$\blacksquare$}}&&&&&&\\
        \bottomrule[1.5pt]
      \end{tabularx}}
    {\caption[Abdomen CT-CT]{Abdomen CT-CT}
      \label{tabAbdomenCTCT}}
    \hfill%
    \ttabbox%
    {\begin{tabularx}{0.49\textwidth}{rCCCCCc}
        \toprule[1.5pt]
        & DSC$\uparrow$ & DSC9$\uparrow$ & HD95$\downarrow$ & SDLogJ$\downarrow$ & RT$\downarrow$ & Rank$\uparrow$ \\
        \midrule
        Initial & 0.33 & 0.22 & 48.65 & & & \\
        \midrule
        ConvexAdam \textcolor{ConvexAdam}{\scalebox{1.25}{$\blacksquare$}} & 0.75 & 0.73 & 24.92 & 0.09 & 1.30 & \colorbox{gold}{0.82$\pm$0.01}\\
        corrField \textcolor{corrField}{\scalebox{1.25}{$\blacksquare$}} & 0.76 & 0.73 & 23.35 & 0.10 & 2.13 & \colorbox{silver}{0.81$\pm$0.02}\\
        LapIRN \textcolor{LapIRN}{\scalebox{1.25}{$\blacksquare$}} & 0.76 & 0.69 & 22.81 & 0.12 & 1.50 & \colorbox{bronze}{0.77$\pm$0.03}\\
        PIMed \textcolor{PIMed}{\scalebox{1.25}{$\blacksquare$}} & 0.78 & 0.68 & 21.99 & 0.07 & 59.2 & 0.75$\pm$0.02\\
        MEVIS \textcolor{MEVIS}{\scalebox{1.25}{$\blacksquare$}} & 0.71 & 0.65 & 27.94 & 0.15 & 14.7 & 0.67$\pm$0.02\\
        Driver \textcolor{Driver}{\scalebox{1.25}{$\blacksquare$}} & 0.76 & 0.55 & 27.02 & 0.13 & 1.95 & 0.63$\pm$0.03\\
        NiftyReg \textcolor{NiftyReg}{\scalebox{1.25}{$\blacksquare$}} & 0.65 & 0.55 & 33.09 & 0.12 & 11.0 & 0.55$\pm$0.02\\
        LaTIM \textcolor{Jaouen}{\scalebox{1.25}{$\blacksquare$}} & 0.54 & 0.49 & 41.17 & 0.13 & & 0.39$\pm$0.03\\
        Winter \textcolor{Winter}{\scalebox{1.25}{$\blacksquare$}} & 0.55 & 0.41 & 35.51 & 0.85 & 2.79 & 0.31$\pm$0.03\\
        Imperial \textcolor{Imperial}{\scalebox{1.25}{$\blacksquare$}} & 0.51 & 0.41 & 48.60 & 0.11 & 278 & 0.30$\pm$0.02\\
        Multi-brain \textcolor{Brudfors}{\scalebox{1.25}{$\blacksquare$}} & 0.54 & 0.44 & 38.21 & 0.48 & & 0.30$\pm$0.02\\
        \bottomrule[1.5pt]
      \end{tabularx}}
    {\caption[Abdomen MR CT]{Abdomen MR-CT}
      \label{tabAbdomenMRCT}}
  \end{floatrow}
  \vspace*{1em}
  \begin{floatrow}[2]
    \ttabbox%
        {\begin{tabularx}{0.49\textwidth}{rCCCCCc}
                \toprule[1.5pt]
        & DSC$\uparrow$ & DSC30$\uparrow$ & HD95$\downarrow$ & SDLogJ$\downarrow$ & RT$\downarrow$ & Rank$\uparrow$ \\
        \midrule
        Initial & 0.56 & 0.27 & 3.86 & & & \\
        \midrule
        LapIRN \textcolor{LapIRN}{\scalebox{1.25}{$\blacksquare$}} & 0.82 & 0.66 & 1.67 & 0.07 & 1.21 & \colorbox{gold}{0.92$\pm$0.01}\\
        ConvexAdam \textcolor{ConvexAdam}{\scalebox{1.25}{$\blacksquare$}} & 0.81 & 0.64 & 1.63 & 0.07 & 3.10 & \colorbox{silver}{0.82$\pm$0.01}\\
        lWM \textcolor{IWM}{\scalebox{1.25}{$\blacksquare$}} & 0.79 & 0.61 & 1.84 & 0.05 & 2.55 & \colorbox{bronze}{0.79$\pm$0.02}\\
        Driver \textcolor{Driver}{\scalebox{1.25}{$\blacksquare$}} & 0.80 & 0.62 & 1.77 & 0.08 & 2.02 & 0.75$\pm$0.02\\
        PIMed \textcolor{PIMed}{\scalebox{1.25}{$\blacksquare$}} & 0.78 & 0.58 & 1.86 & 0.06 & 3.47 & 0.71$\pm$0.02\\
        3Idiots \textcolor{3Idiots}{\scalebox{1.25}{$\blacksquare$}} & 0.80 & 0.63 & 1.82 & 0.08 & 1.46 & 0.70$\pm$0.02\\
        Winter \textcolor{Winter}{\scalebox{1.25}{$\blacksquare$}} & 0.77 & 0.57 & 2.16 & 0.08 & 2.56 & 0.55$\pm$0.02\\
        MEVIS \textcolor{MEVIS}{\scalebox{1.25}{$\blacksquare$}} & 0.77 & 0.57 & 2.09 & 0.07 & 10.4 & 0.51$\pm$0.02\\
        Multi-brain \textcolor{Brudfors}{\scalebox{1.25}{$\blacksquare$}} & 0.78 & 0.59 & 1.92 & 0.57 & & 0.38$\pm$0.02\\
        corrField \textcolor{corrField}{\scalebox{1.25}{$\blacksquare$}} & 0.74 & 0.51 & 2.36 & 0.08 & 5.14 & 0.37$\pm$0.02\\
        Thorley \textcolor{Thorley}{\scalebox{1.25}{$\blacksquare$}} & 0.77 & 0.60 & 2.21 & 0.31 & & 0.37$\pm$0.02\\
        NiftyReg \textcolor{NiftyReg}{\scalebox{1.25}{$\blacksquare$}} & 0.73 & 0.51 & 2.37 & 0.06 & 5.00 & 0.36$\pm$0.01\\
        Bailiang \textcolor{Bailiang}{\scalebox{1.25}{$\blacksquare$}} & 0.67 & 0.42 & 2.74 & 0.04 & 1.38 & 0.33$\pm$0.00\\
        LaTIM \textcolor{Jaouen}{\scalebox{1.25}{$\blacksquare$}} & 0.74 & 0.52 & 2.31 & 0.08 & & 0.32$\pm$0.01\\
        Imperial \textcolor{Imperial}{\scalebox{1.25}{$\blacksquare$}} & 0.76 & 0.57 & 2.43 & 0.19 & 2610 & 0.29$\pm$0.01\\
        \bottomrule[1.5pt]
      \end{tabularx}}
    {\caption[OASIS]{OASIS}
      \label{tabOASIS}}
    \hfill%
    \ttabbox%
    {\begin{tabularx}{0.49\textwidth}{rCCCCc}
        \toprule[1.5pt]
        & TRE$\downarrow$ & TRE30$\downarrow$ & SDLogJ$\downarrow$ & RT$\downarrow$ & Rank$\uparrow$ \\
        \midrule
        Initial & 10.24 & 16.80 & & & \\
        \midrule
        corrField \textcolor{corrField}{\scalebox{1.25}{$\blacksquare$}} & 1.75 & 2.48 & 0.05 & 2.91 & \colorbox{gold}{0.87$\pm$0.01}\\
        ConvexAdam \textcolor{ConvexAdam}{\scalebox{1.25}{$\blacksquare$}} & 1.79 & 2.70 & 0.06 & 1.82 & \colorbox{silver}{0.81$\pm$0.01}\\
        MEVIS \textcolor{MEVIS}{\scalebox{1.25}{$\blacksquare$}} & 1.68 & 2.37 & 0.08 & 95.4 & \colorbox{bronze}{0.78$\pm$0.01}\\
        LapIRN \textcolor{LapIRN}{\scalebox{1.25}{$\blacksquare$}} & 1.98 & 2.95 & 0.06 & 10.3 & 0.73$\pm$0.02\\
        PDD-Net \textcolor{PDD-Net}{\scalebox{1.25}{$\blacksquare$}} & 2.46 & 3.81 & 0.04 & 4.22 & 0.62$\pm$0.02\\
        LaTIM \textcolor{Jaouen}{\scalebox{1.25}{$\blacksquare$}} & 1.83 & 2.50 & 0.05 & & 0.62$\pm$0.01\\
        Lifshitz \textcolor{Lifshitz}{\scalebox{1.25}{$\blacksquare$}} & 2.26 & 3.01 & 0.07 & 2.90 & 0.61$\pm$0.02\\
        Imperial \textcolor{Imperial}{\scalebox{1.25}{$\blacksquare$}} & 1.81 & 2.54 & 0.11 & 300 & 0.57$\pm$0.01\\
        PIMed \textcolor{PIMed}{\scalebox{1.25}{$\blacksquare$}} & 2.34 & 3.27 & 0.04 & 623 & 0.55$\pm$0.02\\
        NiftyReg \textcolor{NiftyReg}{\scalebox{1.25}{$\blacksquare$}} & 2.70 & 5.28 & 0.10 & 42.2 & 0.51$\pm$0.02\\
        Driver \textcolor{Driver}{\scalebox{1.25}{$\blacksquare$}} & 2.66 & 3.50 & 0.10 & 2.66 & 0.44$\pm$0.02\\
        Winter \textcolor{Winter}{\scalebox{1.25}{$\blacksquare$}} & 7.41 & 10.11 & 0.09 & 12.0 & 0.40$\pm$0.02\\
        Epicure \textcolor{Fourcade}{\scalebox{1.25}{$\blacksquare$}} & 6.55 & 10.29 & 0.07 & & 0.29$\pm$0.02\\
        Multi-brain \textcolor{Brudfors}{\scalebox{1.25}{$\blacksquare$}} & 6.61 & 8.75 & 0.08 & & 0.27$\pm$0.01\\
        Gunnarsson \textcolor{Gunnarsson}{\scalebox{1.25}{$\blacksquare$}} & 9.00 & 11.27 & 0.12 & 30.9 & 0.21$\pm$0.00\\
        \bottomrule[1.5pt]
      \end{tabularx}}
    {\caption[Lung CT]{Lung CT}
      \label{tablungCTCT}}
  \end{floatrow}
  \vspace*{1em}
  \begin{floatrow}[1]
     \ttabbox%
    {\begin{tabularx}{1.02\textwidth}{rCCCCCC|C|CC}
        \toprule[1.5pt]
        & CuRIOUS & Hippocampus MR & Abdomen CT-CT & Abdomen MR-CT & OASIS & Lung\hspace{25pt} CT & Overall & Intra-\hspace{10pt}Patient & Inter-\hspace{10pt}Patient \\
        \midrule
        ConvexAdam \textcolor{ConvexAdam}{\scalebox{1.25}{$\blacksquare$}} & \colorbox{bronze}{0.77} & \colorbox{bronze}{0.75} & \colorbox{gold}{0.94} & \colorbox{gold}{0.82} & \colorbox{silver}{0.82} & \colorbox{silver}{0.81} & \colorbox{gold}{0.82} & \colorbox{silver}{0.80} & \colorbox{silver}{0.83} \\
        LapIRN \textcolor{LapIRN}{\scalebox{1.25}{$\blacksquare$}} & 0.49 & \colorbox{gold}{0.93} & \colorbox{silver}{0.82} & \colorbox{bronze}{0.77} & \colorbox{gold}{0.92} & 0.73 & \colorbox{silver}{0.76} & \colorbox{bronze}{0.65} & \colorbox{gold}{0.89} \\
        MEVIS \textcolor{MEVIS}{\scalebox{1.25}{$\blacksquare$}} & 0.42 & \colorbox{silver}{0.78} & 0.60 & 0.67 & 0.51 & \colorbox{bronze}{0.78} & \colorbox{bronze}{0.61} & 0.61 & \colorbox{bronze}{0.62} \\
        corrField \textcolor{corrField}{\scalebox{1.25}{$\blacksquare$}} & \colorbox{gold}{0.85} & 0.34 & 0.53 & \colorbox{silver}{0.81} & 0.37 & \colorbox{gold}{0.87} & 0.59 & \colorbox{gold}{0.84} & 0.41 \\
        NiftyReg \textcolor{NiftyReg}{\scalebox{1.25}{$\blacksquare$}} & 0.56 & 0.37 & 0.36 & 0.55 & 0.36 & 0.51 & 0.44 & 0.54 & 0.36 \\
        PIMed \textcolor{PIMed}{\scalebox{1.25}{$\blacksquare$}} &  &  & 0.49 & 0.75 & 0.71 & 0.55 & 0.35 & 0.39 & 0.33 \\
        PDD-Net \textcolor{PDD-Net}{\scalebox{1.25}{$\blacksquare$}} & \colorbox{silver}{0.83} & 0.58 & 0.44 &  &  & 0.62 & 0.34 & 0.37 & 0.32 \\
        Gunnarsson \textcolor{Gunnarsson}{\scalebox{1.25}{$\blacksquare$}} & 0.19 & 0.25 & 0.33 &  &  & 0.21 & 0.19 & 0.16 & 0.22 \\
        \bottomrule[1.5pt]
    \end{tabularx}}
    {\caption[Overall Rank]{Overall rank scores of methods submitted to four or more tasks.}
    \label{tabResults}
      \label{ranks}}
  \end{floatrow}
\end{table*}%

\begin{figure*}
\RawFloats
   \centering
    \includegraphics[height=0.92\textheight,width=0.95\textwidth]{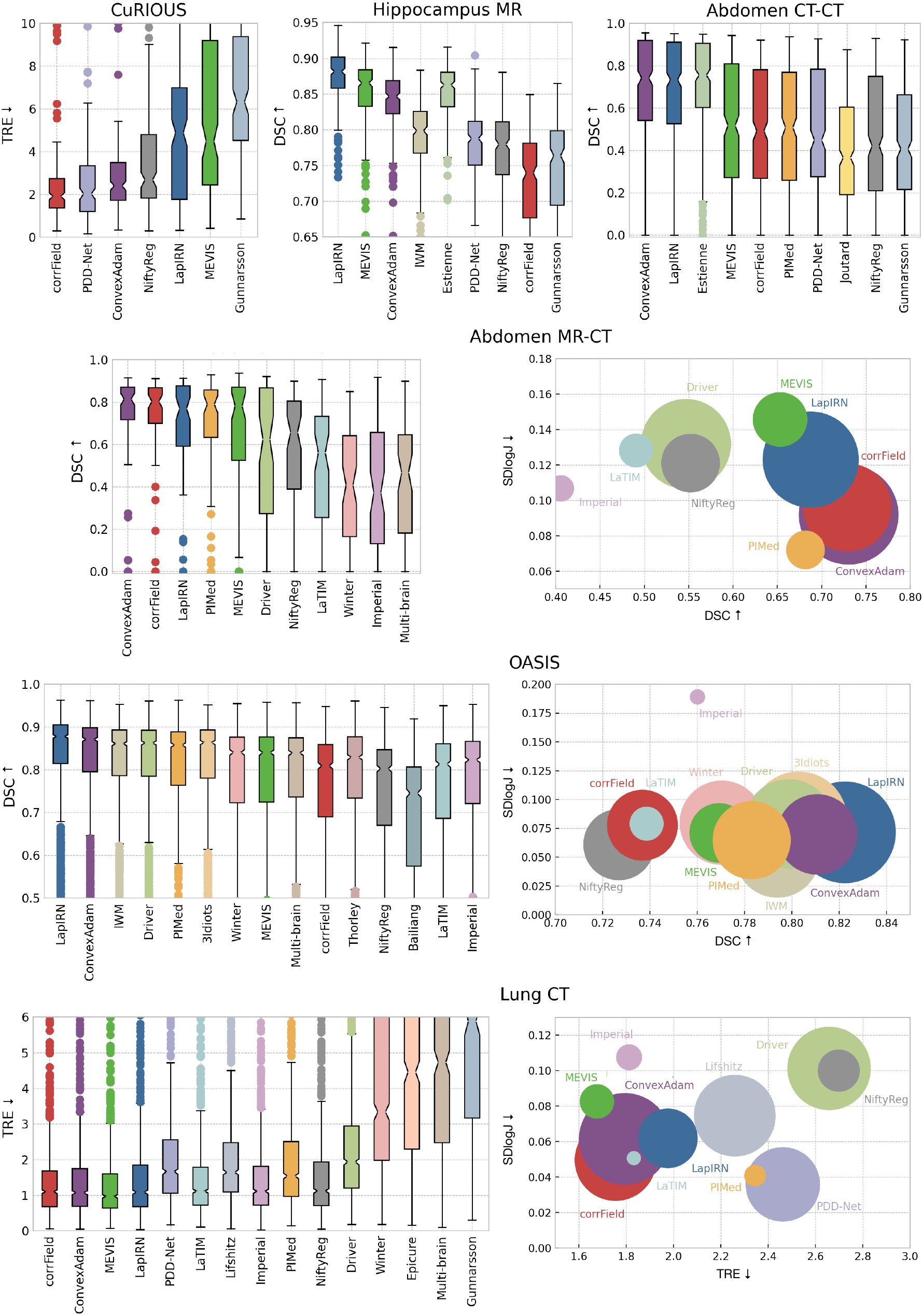}
    \caption{Boxplots and (selected) bubble charts visualising the results for the six challenge tasks. While the boxplots show the main accuracy metric (DSC and TRE, respectively), the bubble charts combine the accuracy, smoothness and runtime metric (a larger bubble means a faster runtime). Arrows ($\uparrow$,$\downarrow$) indicate the favourable direction of metrics. Comparison methods are color coded: ConvexAdam~\textcolor{ConvexAdam}{\scalebox{1.25}{$\blacksquare$}}, LapIRN~\textcolor{LapIRN}{\scalebox{1.25}{$\blacksquare$}}, MEVIS~\textcolor{MEVIS}{\scalebox{1.25}{$\blacksquare$}}, corrField~\textcolor{corrField}{\scalebox{1.25}{$\blacksquare$}}, NiftyReg~\textcolor{NiftyReg}{\scalebox{1.25}{$\blacksquare$}}, PDD-Net~\textcolor{PDD-Net}{\scalebox{1.25}{$\blacksquare$}}, PIMed~\textcolor{PIMed}{\scalebox{1.25}{$\blacksquare$}}, Gunnarsson~\textcolor{Gunnarsson}{\scalebox{1.25}{$\blacksquare$}}, lWM~\textcolor{IWM}{\scalebox{1.25}{$\blacksquare$}}, Estienne~\textcolor{Estienne}{\scalebox{1.25}{$\blacksquare$}}, Joutard~\textcolor{Joutard}{\scalebox{1.25}{$\blacksquare$}}, Driver~\textcolor{Driver}{\scalebox{1.25}{$\blacksquare$}}, LaTIM~\textcolor{Jaouen}{\scalebox{1.25}{$\blacksquare$}}, Winter~\textcolor{Winter}{\scalebox{1.25}{$\blacksquare$}}, Imperial~\textcolor{Imperial}{\scalebox{1.25}{$\blacksquare$}}, Multi-brain~\textcolor{Brudfors}{\scalebox{1.25}{$\blacksquare$}}, 3Idiots~\textcolor{3Idiots}{\scalebox{1.25}{$\blacksquare$}}, Thorley~\textcolor{Thorley}{\scalebox{1.25}{$\blacksquare$}}, Bailiang~\textcolor{Bailiang}{\scalebox{1.25}{$\blacksquare$}}, Epicure~\textcolor{Fourcade}{\scalebox{1.25}{$\blacksquare$}}, and Lifshitz~\textcolor{Lifshitz}{\scalebox{1.25}{$\blacksquare$}}. Methods are sorted according to final rank scores.}
    \label{fig:boxplotBubbleCharts}
\end{figure*}
  
\subsection{Challenge Outcome}
In this section, we will first present each task separately and subsequently the eight methods that are included in the overall ranking. Tables \ref{tabCuRIOUS} to \ref{tablungCTCT} give the numerical results and the scores for each algorithm for each task averaged over the anatomical structures/landmarks and number of scan pairs that were registered for that task. The algorithms are listed in order of their final placement per task. Standard deviations of final rank scores are calculated using \removed{leave-one-out bootstrapping}\revised{jackknife resampling\cite{tukey1958bias}}. Fig. \ref{fig:boxplotBubbleCharts} shows boxplots illustrating the distribution of the accuracy (TRE and Dice) of the different methods for each task. Furthermore, for selected task (Abdomen MR-CT, OASIS, and Lung CT), a bubble chart combines the accuracy, smoothness, and runtime metric.

\paragraph*{\textbf{CuRIOUS}}
Four methods were submitted to this task in addition to the three baseline methods. For two of these methods, some cases caused negative outliers and the average TRE was worse than the initial TRE (c.f. Table \ref{tabCuRIOUS}). Only the registration of the two baseline methods corrField and PDD-Net as well as the ConvexAdam method led to a considerable reduction in TRE from 6.38~mm to 2.84~mm, 3.08~mm, and 3.31~mm, respectively.

\paragraph*{\textbf{Hippocampus MR}}
In this task, all algorithms consistently performed very well (median Dice~$>$~0.7). Nevertheless, there is a performance gap between algorithms using label supervision (LapIRN, MEVIS, ConvexAdam, and Estienne) and unsupervised methods (NiftyReg, PDD-Net and corrField). However, despite label-supervision, the methods of IWM and Gunnarsson perform comparably to unsupervised methods. This is the only task that enabled sub-second runtimes.

\paragraph*{\textbf{Abdomen CT-CT}}
In this task, a clear three-way partition of the algorithms appears. The methods of Estienne, LapIRN, and ConvexAdam achieved a Dice Score of 0.67-0.69 across the eight individual organs and thus at least a 0.2 higher Dice Score then all other participants. The midfield includes the unsupervised methods MEVIS, corrField, and PDD-Net and the supervised method PIMed which achieve a Dice Score of 0.49-0.51. The final group is formed by the methods Joutard, NiftyReg and Gunnarsson with a Dice Score of 0.40-0.45. This structure can also be found in the other accuracy measures DSC30 and HD95. All methods, apart from NiftyReg and Gunnarsson, have a runtime of fewer than 10 seconds.

\begin{figure*}[t]
\centering
\includegraphics[width=0.7\textwidth]{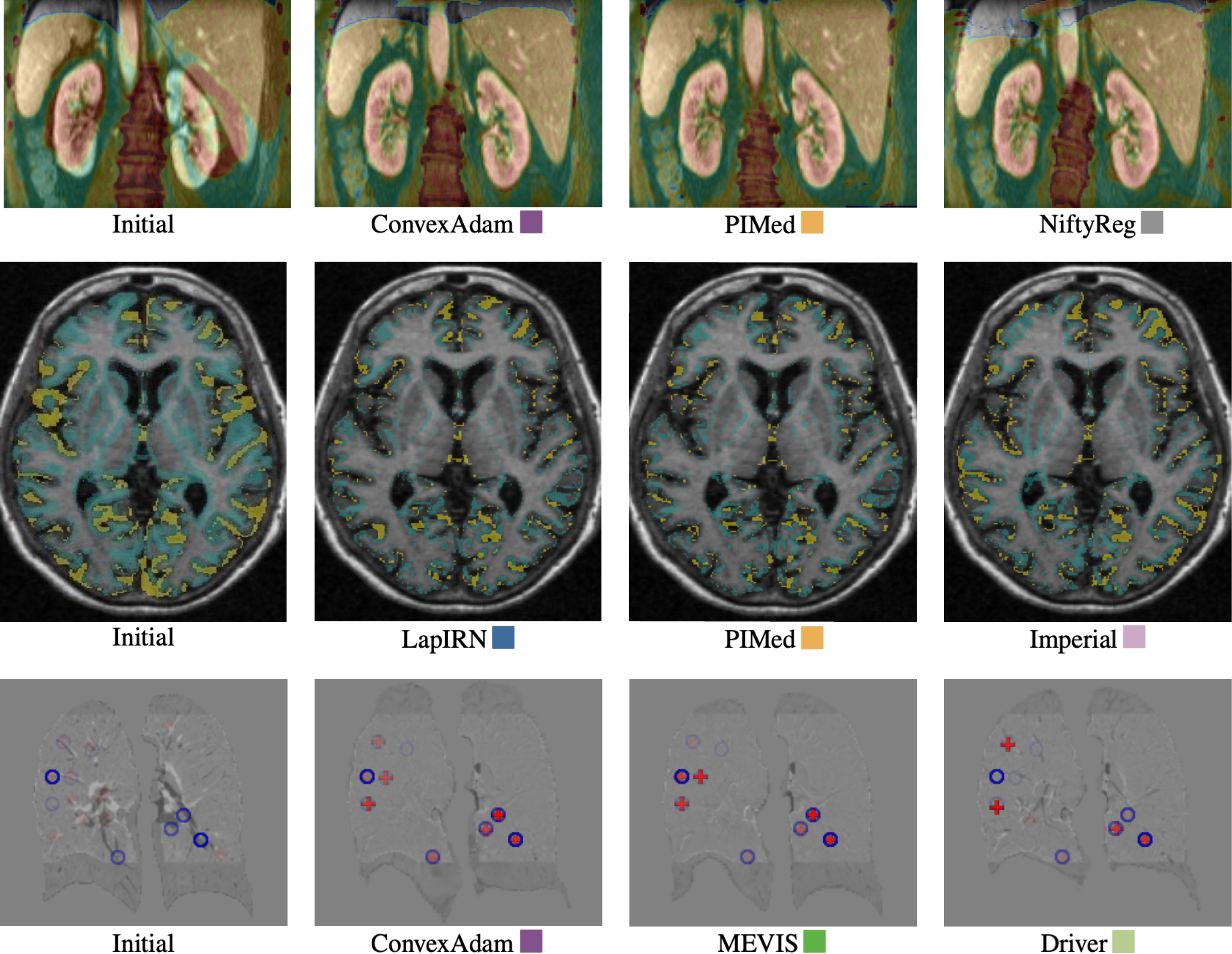}
\caption{Exemplary qualitative results for selected methods and tasks. Top row: Overlay of coronal abdominal MR (gray) and warped CT (color) slices. Middle row: False-negative (green) and false-positive (yellow) voxels of propagated segmentation labels on transversal slices of the OASIS dataset. Bottom row: Coronal slices of difference images between exhale and warped inhale lung CT scans (including exhale (blue circle) and warped inhale (red cross) landmarks).}
\label{fig:ResultsImages}
\end{figure*}

\paragraph*{\textbf{Abdomen MR-CT}}\label{sec:AbdomenMRCTresults}
In the abdominal MR-CT task, the algorithms can also be divided into three groups based on the median Dice Score (c.f. Fig.~\ref{fig:boxplotBubbleCharts}). The leading group can be further divided into the algorithms that achieve a similar Dice Score on the segmentations provided in the training as on the nine unknown organ segmentations (ConvexAdam and corrField) and those that show a performance loss on the nine unknown organs (LapIRN, PIMed, MEVIS). This division is also reflected in the variance of the achieved Dice Scores. In respect of runtime, PIMed stands out in this task with a runtime of approximatly one minute. In Fig.~\ref{fig:ResultsImages}, exemplary qualitative registration results are shown.

\paragraph*{\textbf{OASIS}}
The OASIS inter-subject brain task attracted the most learning-based solutions. The results are summarised in Table~\ref{tabOASIS} and visualised in Fig.~\ref{fig:boxplotBubbleCharts} showing that most of these methods achieve very similar results in terms of Dice Score for the cases with the highest scores (Dice of 80-90\%). The differences are primarily in the more difficult cases and thus in the DSC30 score, where the LapIRN, ConvexAdam, and the methods of Driver and 3Idiots methods perform slightly better than for example PIMed and Winter. The conventional methods of MEVIS and corrField achieve mid-ranked accuracies and have a higher runtime. Fig.~\ref{fig:ResultsImages} shows an example transversal slice of the fixed image overlayed with the false-negative segmented voxels (green) and false-positive segmented voxels (yellow) for initial moving segmentation and the propagated segmentations by the methods of Imperial, PIMed, and LapIRN. All methods were able to align the small structures of the brain with only very small visible differences.

\paragraph*{\textbf{Lung CT}}
This task was carried out in both years because in 2020 only the MEVIS, which uses automatically computed keypoints as additional metric, achieved a TRE of less than 2mm (1.72mm), while other teams performed considerably worse (e.g. LapIRN 3.24mm and PDD-Net 2.46mm). In 2021, keypoint correspondences were provided for training and the submissions improved, with six teams (corrField, ConvexAdam, MEVIS, LapIRN, LaTIM, Liftschitz) achieving a TRE of less than 2mm. Compared to the other tasks, the runtime in the lung CT task is considerably longer for several algorithms due to the additional time needed to compute keypoints or perform instance optimisation. Fig.~\ref{fig:ResultsImages} visualises the difference images of an example coronal slices for the methods of Driver, ConvexAdam, and MEVIS overlayed with manual landmarks.

\paragraph*{\textbf{Overall Ranking}}
Table~\ref{ranks} gives the overall rank scores of the eight methods submitted to four or more tasks. Additionally, we separately listed the scores for inter- and intra-patient registration tasks. ConvexAdam was among the top three on each task (winning Abdomen CT-CT and Abdomen MR-CT) and ranked first overall. The GPU-acceleration brings down computation cost of this optimisation-based method to a few seconds for 3D registration and that is why it consistently achieves high scores for the run time in addition to the very good quality scores. LapIRN reached the overall second rank and yielded the best result for Hippocampus MR and OASIS. This demonstrates that a well-designed convolutional feed-forward network (instance optimisation was used only for CuRIOUS and Lung CT) can outperform conventional approaches in particular for inter-patient tasks. MEVIS achieved the third place overall, with top ranks in particular for Lung CT and Hippocampus MR based on a combination of NGF metric, curvature regularisation, and L-BFGS optimisation with additional learning components only employed for the brain task. CorrField uses no label supervision at all, but relies on highly optimised graph-based registration, and comes fourth overall winning two individual tasks: CuRIOUS and LungCT. It is the best method for intra-patient registration. PIMed's method achieves strong performance on Abdomen MR-CT and OASIS and generalises well to Abdomen CT-CT.

\subsection{Additional Experiments}{\label{sec:AdditionalExperiments}}
\paragraph*{\textbf{Label Bias and Unsupervised Registration}} When evaluating the influence of supervision with anatomical labels, we found a clear distinction between intra- (Abdomen MR-CT) and inter-patient registration (Abdomen CT-CT, Hippocampus and OASIS), see Table \ref{tabResults}. The former shows nearly no advantage of including such information and it is therefore possible to avoid a risk of overfitting towards certain anatomies. The latter, however, shows a clear deterioration in accuracy when excluding structures from training that are used for evaluation. CorrField (unsupervised) achieves the highest scores for intra-patient registration trails nearly all learning-based methods on the remaining inter-patient tasks. LapIRN trained without Dice loss (i.e. without anatomical knowledge) improves upon those results and achieves very strong results for OASIS and Abdomen CT-CT. This indicates that a large training database and an advanced deep learning architecture may narrow the gap between supervised and unsupervised approaches. We evaluated ConvexAdam for Abdomen CT-CT in three settings, each time evaluating on 8 test labels: 1) all 13 labels in training (DSC=69\%), 2) 4 labels in training (DSC=55\%) and 3) no labels in training (DSC=45\%). This shows that partial supervision clearly leads to improvement of those identical anatomies but can also help to align nearby structures: aligning the esophagus which was excluded as label from training improved by 16\% points (likely through the guidance of liver and aorta) and pancreas overlap was increased by 12\% points (possibly by including portal vein and adrenal gland). As mentioned in Sec. \ref{sec:AbdomenMRCTresults} training on 4 and evaluating on 9 abdominal organs for MR-CT fusion results in a moderate performance gap between supervised and unsupervised methods.
\paragraph*{\textbf{Transferability}} We were able to show that the three best methods of the lung registration task also perform very well on the DIRLab dataset (MEVIS 1.22~mm, ConvexAdam 1.31~mm, and corrField 1.34~mm) without further hyperparameter adaptations. Since the inspiration and expiration images of the DIRLab dataset are extracted from a 4DCT dataset with shallow breathing, the registration task is probably easier than the Learn2Reg lung CT task. This might explain the lower TRE values on the DIRLab dataset compared to the Learn2Reg lung task (improved TRE of 0.46~mm, 0.48~mm, and 0.41~mm for MEVIS, ConvexAdam, and corrField, respectively). Due to the preprocessing and the reduced resolutions, the Learn2Reg methods achieve slightly worse results than state-of-the-art methods evaluated on the DIRLab dataset. For example, the method of MEVIS as part of their complete registration pipeline and applied to the original images reaches a TRE of 0.94~mm~\cite{ruhaak2017estimation}. LapIRN achieves similar results on both datasets (Learn2Reg lung CT 1.98~mm and DIRLab 1.98~mm) showing that the best deep-learning-based methods can also be successfully applied to other datasets without retraining.

\section{Discussion}
\paragraph*{\textbf{Reducing Entry Barriers}} By pre-processing each dataset to the same dimensions and isotropic resolution and providing anatomical annotations for training data wide participation was achieved from research groups across the world. The OASIS inter-subject brain task attracted the most learning-based solutions, which highlights the importance of large, labelled training datasets for deep-learning registration and mirrors the focus of recent research. Lung CT intra-patient registration was addressed by the same number but more diverse set of methods, including conventional, fully deep-learning-based, and hybrid approaches. Some aspects of medical image registration, including affine or rigid pre-alignment, dealing with differences in field-of-view of voxel resolutions, and the processing of very high-resolution scans have been omitted due to our challenge design and could be addressed in future.

\paragraph*{\textbf{Task specific results}}
In general, it is difficult to find the exact reasons why one or the other method performed better or worse in the various tasks. Nevertheless, there are some relevant patterns that can be identified. In the CuRIOUS task, the three methods using a dense discretised displacement correlation (ConvexAdam, corrField and PDD-Net) cope best with the difference in the field of view of the input images. In the case of the Hippocampus MR task, the learning-based methods perform considerably better. This can be explained by the fact that the structures used for the evaluation were also available in the training data set, so that the learning-based methods were specialised in the alignment of these structures during the training. A similar result was observed on the OASIS and Abdomen CT-CT task. The OASIS dataset has already been used in the past in various training-test splits by several groups to develop and test the registration algorithms, so that consistently good results were to be expected and which became true for both deep-learning based and conventional methods. On the Abdomen CT-CT dataset, it is difficult to explain the large performance difference of a nearly 0.2 higher Dice. A successful strategy for inter patient registration can be identified in the ConvexAdam method. Instead of using the segmentations directly in the training of a registration network, a segmentation network is trained. This is used to first generate the segmentations on new data and then to utilise them in the cost function of the optimisation-based registration. In the Abdomen MR-CT task, we found that using a \removed{dice}\revised{Dice} loss for certain structures can lead to overfitting on these structures and therefore the registration network might not registering other structures as well. Furthermore, it has been shown that a multimodal distance metric, as used by most participants, is essential. Successful strategies for lung registration seem to be the use of keypoints and the combination of deep learning registration + instance optimisation. Gunnarsson's learning-based method performs worse in comparison, this is most likely due to the fact that a common network was trained for the Lung CT, Abdomen CT-CT and Hippocampus MR tasks showing that task-specific solutions might be beneficial. Nevertheless, this result shows that a registration network is capable of solving very different tasks at the same time.

\paragraph*{\textbf{Comparison of Learning- vs Optimisation-based Registration}} We argue that Learn2Reg has helped to demystify common beliefs of fundamental differences between learning- and optimisation registration. First and foremost, there is virtually no difference in computational speed. GPU-acceleration brings down computation cost of optimisation-based methods to a few seconds for 3D registration, i.e. the extraction of features using CNNs often outweighs optimisation times. Furthermore, we see a clear trend that learning on segmentation labels is primarily beneficial for inter-subject registration. For Abdomen CT-CT for instance large improvements of 20\%points in Dice overlap compared to previous work~\cite{xu2016evaluation} could be achieved using Dice losses. All three highest ranked approaches employ a combination of DL and optimisation: LapIRN primarily uses a deep network, but add instance optimisation for Lung CT, MEVIS mainly use conventional optimisation but a DL network for Hippocampus MR, and ConvexAdam combines discrete optimisation with UNet-based semantic features for inter-patient tasks. Our current challenge design did not consider any computational constraints (GPU memory, runtime on CPU), which might limit the practical impact for some applications and should be considered in future studies.

\paragraph*{\textbf{Algorithmic Design Choices}} There are no direct ablation studies possible for the used architectures and loss functions since each method differs in multiple aspects (see Table \ref{tab:overviewParticipants}), but some general trends are visible nonetheless. Most approaches use a combination of contrast-invariant intensity metrics (LNCC, NGF and MIND) as well as a Dice loss for tasks where anatomical labels are available. To address larger motion (all tasks expect brain) DL registration methods employ multi-scale (and residual) architectures, multiple warps or often dense correlation layers. Two-stream approaches that process both input scans independently are commonplace to deal with multimodality or contrast variations.

\paragraph*{\textbf{Comparison to Baselines}} We evaluated two conventional methods, NiftyReg~\cite{modat2010fast} and corrField~\cite{heinrich2015estimating} (using the GPU implementation of~\cite{hansen2021graphregnet}), and two learning-based approaches, PDD-Net~\cite{heinrich2019closing} and the original VoxelMorph~\cite{balakrishnan2019voxelmorph} as baselines. The latter two were only applied to a subset of tasks. NiftyReg achieves reasonable accuracies but falls behind supervised methods on inter-patient tasks. The original VoxelMorph variant reaches an average Dice overlap of 76.88\%$\pm$2.17~\% for OASIS (7th-10th place based on DSC alone) and a TRE of 7.51$\pm$3.43~mm for lung CT (13th place). When trained on a large additional lung dataset~\cite{hering2021cnn} a TRE of 1.71$\pm$2.86~mm was achieved for the additional DIRLAB lung experiment for which the best performing methods in this challenge achieved 1.3~mm. PDD-Net achieved a second rank for CuRIOUS and fifth place for Lung CT. CorrField achieved the best scores overall for CuRIOUS and LungCT and second place for Abdomen MR-CT, making it stand out as the best performing intra-patient approach (without supervision). This demonstrates that conventional methods are still very competitive for datasets without strong label supervision.

\paragraph*{\textbf{Plausibility of Transformations}} We analysed the smoothness of transformations with respect to the log-standard deviation of Jacobian determinants for all experiments. While this measure is far from perfect, it enabled a ranking of different solutions to the inherently ambiguous nonlinear registration task that may achieve similar accuracy with large differences in complexity (the common assumption being: the smoother transform is then preferable). As visualised in Fig. \ref{fig:boxplotBubbleCharts} there is a tendency that more accurate solutions are also smoother, which indicates that enforcing regularity is an effective means of avoiding overfitting and improves robustness. Some notable exceptions can be found for lung CT, where Imperial appears to suffer from too low regularisation while PDD-Net and PIMed may have reduced accuracy in exchange for overly smooth fields. A potential explanation for the positive correlation of smoothness and accuracy could be the hypothesis that accurate methods are able to establish strong (correct) correspondences at relevant anatomies and extrapolate as smooth as possible in uncertain areas. That means putting emphasis on either surfaces (e.g. based on segmentation estimates) or geometric keypoints (for lung scans) can be beneficial.

\paragraph*{\textbf{Limitations of the Challenge Design}} We have identified a number of limitations that should be addressed in future studies. First, for computational reasons the training of algorithms was performed offline by participants. This could introduce a bias when additional data is used by certain teams that is not accessible to others and prevents the use of larger datasets that cannot be made public due to privacy concerns. Enabling docker-based training or fine-tuning of models directly at grand-challenge.org would be desirable. Second, the amount of available annotated training data varied across tasks and made in particular intra-patient tasks harder for learning-based approaches. Unfortunately, the problem is that large datasets are often not publicly available and therefore cannot be used in this type of challenge. Decoupling anatomical feature learning from patient-wise optimisation could be a next step, e.g. by providing training data for airway and fissure segmentation for lung CT. The registration accuracy cannot be measured directly but must be evaluated via auxiliary metrics such as the overlap of segmentation masks which disregards the plausibility of correspondences along the surface or within the structure. While this is an inherent problem in evaluating image registration, this issue can be mitigated by generating further manual annotations for certain structures. The provision of all segmentation classes for training that were used for testing is in our opinion the most problematic limitation of this challenge. This was due to the fact, that for 3 out of 4 tasks with segmentation labels these annotations were already publicly available prior to Learn2Reg and we considered it in-transparent to simply not point participants to their availability. We aimed to mitigate the influence of over-fitting towards labelled anatomies by performing additional experiments for partial supervision. And finally, statements about the quality of the registration algorithms can only be generalised to a limited extent, but apply mainly to the selected tasks.

\paragraph*{\textbf{Impact and Clinical Adoption}} With regard to the five-year-old survey on medical image registration by~\cite{viergever2016survey}, we can reflect that the shift from surface- to intensity-based registration has somewhat been reverted with a majority of approaches employing segmentation-based overlap or keypoints as driving force. The establishing of learning-based strategies, including hybrid approaches that decouple semantic feature extraction from optimisation or combine feed-forward networks with instance optimisation, can be seen as an important new trend. To assess the likelihood of adopting registration in clinical practice, we are encouraged to see that a number of previous obstacles have been successfully addressed by the participants. First, robustness against variations in scanner protocol and patient characteristics was shown to be very high for top-ranking methods that tackled both multi-centric MR studies (OASIS) as well as the transferability issue for lung CT. Second, run times have been considerably reduced to a few seconds, which will enable clinicians to interact with algorithmic solutions by adjusting hyper-parameters, e.g. the strength of regularisation in near realtime (this holds only true for DL-based methods if they are either decoupled or trained with conditioning cf.~\cite{mok2021conditional}). Third, it became clear that highly nonrigid transformations are as well solved as rigid alignment, opening up the promise for clinical applications in image-guided surgery/radiotherapy. In fact, it appears as if pre-alignment remains an active problem in particular for DL solutions.

\section{Conclusion}
The Learn2Reg challenge was the first to evaluate a wide-range of methods for various inter- and intra-patient as well as mono- and multimodal medical image registration tasks. The main goal was to provide a standardised benchmark on complementary tasks with clinical impact and a platform for comparison of conventional and learning-based medical image registration methods. We established a low entry barrier for training and validation of 3D registration, which helped us compile results of over 65 individual method submissions from more than 20 unique teams. Although registration is highly dependent on the task, two methods \removed{(ConvexAdam, LapIRN,)}\revised{(ConvexAdam and LapIRN)} and a baseline method (corrField) were shown to work robustly on all tasks with only minor adjustments to the hyperparameters. The submission of MEVIS also works robustly for all tasks. It should be noted, however, that they use a deep-learning-based method for the hippocampal tasks. Furthermore, several teams (Estienne, PIMed, Driver, 3idiots, Multi-brain LaTIM, Lifshitz and Imperial) have submitted tailored solutions to individual tasks and achieve very good results with it. Our additional \textit{Transferability experiment} (c.f. section \ref{sec:AdditionalExperiments}) gives a tentative indication that the conventional methods ConvexAdam, MEVIS, and corrField can be directly applied to new data sets without much loss of accuracy. Furthermore, we demystified the common belief that conventional registration methods have to be much slower than deep-learning-based methods. Nevertheless, with LapIRN a deep-learning-based registration method achieves state-of-the-art registration results within seconds. We could not identify any architecture that was advantageous over others. In \removed{or}\revised{our} experiments, it was found that for deep-learning-based methods using a Dice loss for inter-patient registration is particularly useful and instance optimisation helped increasing the accuracy for intra-patient registration. The results presented in this paper initially apply to the submitted methods on the six data sets used in this challenge. However, they may provide a reference for further research on additional data sets. With the Learn2Reg challenge, we have created a dataset for comparing future registration papers. Furthermore, the dataset has the potential to allow the development of dataset-independent and self-configuring registration methods.

\section*{Acknowledgments}
 We thank Yipeng Hu and Tom Vercauteren who co-organised the first Learn2Reg tutorial at MICCAI 2019 which initiated the challenge. The results published here are in part based upon data generated by the TCGA Research Network: https://www.cancer.gov/tcga. \revised{Supported in part by the German Federal Ministry of Education and Research under grant number 031L0202B.}

\ifCLASSOPTIONcaptionsoff
  \newpage
\fi

\bibliographystyle{IEEEtran}
\bibliography{refs.bib}
\footnotesize
A. Hering is with Fraunhofer MEVIS, Institute for Digital Medicine, 23562 L\"ubeck, Germany (email: alessa.hering@mevis.fraunhofer.de) and also with the Department of Radiology and Nuclear Medicine, Radboud University Medical Center, 6525 GA, Nijmegen, The Netherlands

L. Hansen, H.~Siebert, C.~Großbr\"ohmer, M.P.~Heinrich are with with the Institute of Medical Informatics, Universität zu L\"ubeck, 23562 L\"ubeck, Germany.

T.~C.~W.~Mok and A.~C.~S.~Chung are with the Department of Computer Science and Engineering, The Hong Kong University of Science and Technology, Hong Kong.

S. Häger, A. Lange, S. Kuckertz and S. Heldmann are with the Fraunhofer MEVIS, Institute for Digital Medicine, 23562 L\"ubeck, Germany

W.~Shao and M. Rusu are with the Department of Radiology, Stanford University, Stanford CA 94305, USA.

S. Vesal and G. Sonn are with the Department of Urology, Stanford University, Stanford CA 94305, USA

T.~Estienne is with the Université Paris-Saclay, CentraleSupélec, Mathématiques et Informatique pour la
Complexité et les Systèmes, Inria Saclay, 91190, Gif-sur-Yvette, France and also with the Université Paris-Saclay, Institut Gustave Roussy, Inserm, Radiothérapie Moléculaire et Innovation Thérapeutique, 94800, Villejuif, France.

M.~Vakalopoulou is with the Université Paris-Saclay, CentraleSupélec, Mathématiques et Informatique pour la
Complexité et les Systèmes, Inria Saclay, 91190, Gif-sur-Yvette, France.

L.~Han is with the Department of Radiology and Nuclear Medicine, Radboud University Medical Center, 6525 GA, Nijmegen, The Netherlands.

Y.~Huang is with the School of Automation, Nanjing University of Information Science and Technology, Nanjing 210044, China.

M.Brudfors is with the School of Biomedical Engineering and Imaging Sciences, King’s College London, London, UK.

Y. Balbastre is with the Athinoula A. Martinos Center for Biomedical Imaging, Massachusetts General Hospital, USA and also with the Harvard Medical School, Boston, USA.

S. Joutard and M. Modat are with the King's College London, United Kingdom.

G. Lifshitz and D. Raviv are with the Tel Aviv University.

J. Lv and Q. Li are with the Britton Chance Center for Biomedical Photonics, Wuhan National Laboratory for Optoelectronics-Huazhong University of Science and Technology, Wuhan, Hubei 430074, China and also with the MoE Key Laboratory for Biomedical Photonics, Collaborative Innovation Center for Biomedical Engineering, School of Engineering Sciences, Huazhong University of Science and Technology, Wuhan, Hubei 430074, China.

V. Jaouen and D. Visvikis are with the UMR 1101 LaTIM, IMT Atlantique, Inserm, Brest, France.

C. Fourcade is with the Ecole Centrale de Nantes, LS2N, UMR CNRS 6004, Nantes, 44100, France and Keosys Medical Imaging, Saint Herblain, 44300, France.

M. Rubeaux is with the Keosys Medical Imaging, Saint Herblain, 44300, France.

W. Pan is with the Shenzhen International Graduate School, Tsinghua University, China.

Z. Xu is with the Department of Biomedical Engineering, The Chinese University of Hong Kong, Hong Kong, China.

B. Jian and F. De Benetti are with the Chair for Computer Aided Medical Procedures and Augmented Reality, Technische Universität München, Garching, Germany.

M. Wodzinski is with the AGH University of Science and Technology, Department of Measurement and Electronics, Krakow, Poland and also with the University of Applied Sciences Western Switzerland (HES-SO Valais), Information Systems Institute, Sierre, Switzerland.

N. Gunnarsson and Jens Sjölund are with the Department of Information Technology, Uppsala University, Uppsala, Sweden and also with Elekta Instrument AB, Stockholm, Sweden.

H. Qiu and Z. Li are with the Department of Computing at Imperial College London.

A. Hoopes is with the Athinoula A. Martinos Center for Biomedical Imaging, Massachusetts General Hospital, USA.

I. Reinertsen is with the Dept. Health Research, SINTEF Digital, Trondheim, Norway.

Y. Xiao is with the Western University, London, Canada.

B.Landman and Y. Huo are with the Department of Electrical and Computer Engineering, Vanderbilt University.

N Lessmann and K. Murphey and B van Ginnken are with the Department of Radiology and Nuclear Medicine, Radboud University Medical Center, 6525 GA, Nijmegen, The Netherlands.

A. V. Dalca is with the Computer Science and Artificial Intelligence Lab, MIT, USA, the Martinos Center for Biomedical Imaging, Massachusetts General Hospital, USA and also with the Harvard Medical School, Boston, USA.

\end{document}